\RequirePackage{lineno} 
\documentclass[prl,twocolumn,showpacs,amsmath,amssymb,superscriptaddress,showkeys,secnumarabic,preprintnumbers]{revtex4}

\usepackage{graphicx}
\usepackage{dcolumn}
\usepackage{bm}
\usepackage{amssymb}
\usepackage{color}
\usepackage{wrapfig}
\usepackage{multirow}

\newcommand{\pT}{\mbox{$p_\perp$}}

\newcommand{\eg}{\textrm{e.g.}}

\newcommand{\sqrts}{\mbox{$\sqrt{s}$}}
\newcommand{\sqrtsNN}{\mbox{$\sqrt{s_{NN}}$}}
\newcommand{\jpsi}{\mbox{$J/\psi$}}
\newcommand{\upsi}{\mbox{$\Upsilon$}}
\newcommand{\upsione}{\mbox{$\Upsilon(\textrm{1S})$}}
\newcommand{\upsitwo}{\mbox{$\Upsilon(\textrm{2S})$}}
\newcommand{\upsithree}{\mbox{$\Upsilon(\textrm{3S})$}}
\newcommand{\gev}{\mbox{$\mathrm{GeV}$}}
\newcommand{\mev}{\mbox{$\mathrm{MeV}$}}

\newcommand{\gevcc}{\mbox{$\mathrm{GeV}/c^2$}}
\newcommand{\dEdx}{\mbox{$dE/dx$}}
\newcommand{\pp}{\mbox{$p\!+\!p$}}
\newcommand{\AuAu}{\mbox{Au+Au}}
\newcommand{\PbPb}{\mbox{Pb+Pb}}
\newcommand{\dAu}{\mbox{$d$+Au}}
\newcommand{\Npart}{\mbox{$N_{\mathrm{part}}$}}
\newcommand{\Ncoll}{\mbox{$N_{\mathrm{coll}}$}}
\newcommand{\Tc}{\mbox{$T_c$}}
\newcommand{\epluseminus}{\mbox{$e^+e^-$}}
\newcommand{\invpb}{\mbox{$\mathrm{pb}^{-1}$}}
\newcommand{\bbbar}{\mbox{$b$$\bar{b}$}}
\newcommand{\RAA}{\mbox{$R_{AA}$}}

\newcommand{\RdAu}{\mbox{$R_{d\mathrm{Au}}$}}

\begin{document} 


\title{Suppression of \boldmath{$\Upsilon$} Production in \dAu\ and \AuAu\ Collisions at \protect{$\sqrt{s_{NN}} = 200$} GeV} 

\affiliation{AGH University of Science and Technology, Cracow, Poland}
\affiliation{Argonne National Laboratory, Argonne, Illinois 60439, USA}
\affiliation{University of Birmingham, Birmingham, United Kingdom}
\affiliation{Brookhaven National Laboratory, Upton, New York 11973, USA}
\affiliation{University of California, Berkeley, California 94720, USA}
\affiliation{University of California, Davis, California 95616, USA}
\affiliation{University of California, Los Angeles, California 90095, USA}
\affiliation{Universidade Estadual de Campinas, Sao Paulo, Brazil}
\affiliation{Central China Normal University (HZNU), Wuhan 430079, China}
\affiliation{University of Illinois at Chicago, Chicago, Illinois 60607, USA}
\affiliation{Cracow University of Technology, Cracow, Poland}
\affiliation{Creighton University, Omaha, Nebraska 68178, USA}
\affiliation{Czech Technical University in Prague, FNSPE, Prague, 115 19, Czech Republic}
\affiliation{Nuclear Physics Institute AS CR, 250 68 \v{R}e\v{z}/Prague, Czech Republic}
\affiliation{Frankfurt Institute for Advanced Studies FIAS, Germany}
\affiliation{Institute of Physics, Bhubaneswar 751005, India}
\affiliation{Indian Institute of Technology, Mumbai, India}
\affiliation{Indiana University, Bloomington, Indiana 47408, USA}
\affiliation{Alikhanov Institute for Theoretical and Experimental Physics, Moscow, Russia}
\affiliation{University of Jammu, Jammu 180001, India}
\affiliation{Joint Institute for Nuclear Research, Dubna, 141 980, Russia}
\affiliation{Kent State University, Kent, Ohio 44242, USA}
\affiliation{University of Kentucky, Lexington, Kentucky, 40506-0055, USA}
\affiliation{Korea Institute of Science and Technology Information, Daejeon, Korea}
\affiliation{Institute of Modern Physics, Lanzhou, China}
\affiliation{Lawrence Berkeley National Laboratory, Berkeley, California 94720, USA}
\affiliation{Massachusetts Institute of Technology, Cambridge, MA 02139-4307, USA}
\affiliation{Max-Planck-Institut f\"ur Physik, Munich, Germany}
\affiliation{Michigan State University, East Lansing, Michigan 48824, USA}
\affiliation{Moscow Engineering Physics Institute, Moscow Russia}
\affiliation{National Institute of Science Education and Research, Bhubaneswar 751005, India}
\affiliation{Ohio State University, Columbus, Ohio 43210, USA}
\affiliation{Old Dominion University, Norfolk, VA, 23529, USA}
\affiliation{Institute of Nuclear Physics PAN, Cracow, Poland}
\affiliation{Panjab University, Chandigarh 160014, India}
\affiliation{Pennsylvania State University, University Park, Pennsylvania 16802, USA}
\affiliation{Institute of High Energy Physics, Protvino, Russia}
\affiliation{Purdue University, West Lafayette, Indiana 47907, USA}
\affiliation{Pusan National University, Pusan, Republic of Korea}
\affiliation{University of Rajasthan, Jaipur 302004, India}
\affiliation{Rice University, Houston, Texas 77251, USA}
\affiliation{Universidade de Sao Paulo, Sao Paulo, Brazil}
\affiliation{University of Science \& Technology of China, Hefei 230026, China}
\affiliation{Shandong University, Jinan, Shandong 250100, China}
\affiliation{Shanghai Institute of Applied Physics, Shanghai 201800, China}
\affiliation{SUBATECH, Nantes, France}
\affiliation{Temple University, Philadelphia, Pennsylvania, 19122, USA}
\affiliation{Texas A\&M University, College Station, Texas 77843, USA}
\affiliation{University of Texas, Austin, Texas 78712, USA}
\affiliation{University of Houston, Houston, TX, 77204, USA}
\affiliation{Tsinghua University, Beijing 100084, China}
\affiliation{United States Naval Academy, Annapolis, MD 21402, USA}
\affiliation{Valparaiso University, Valparaiso, Indiana 46383, USA}
\affiliation{Variable Energy Cyclotron Centre, Kolkata 700064, India}
\affiliation{Warsaw University of Technology, Warsaw, Poland}
\affiliation{University of Washington, Seattle, Washington 98195, USA}
\affiliation{Yale University, New Haven, Connecticut 06520, USA}
\affiliation{University of Zagreb, Zagreb, HR-10002, Croatia}

\author{L.~Adamczyk}\affiliation{AGH University of Science and Technology, Cracow, Poland}
\author{J.~K.~Adkins}\affiliation{University of Kentucky, Lexington, Kentucky, 40506-0055, USA}
\author{G.~Agakishiev}\affiliation{Joint Institute for Nuclear Research, Dubna, 141 980, Russia}
\author{M.~M.~Aggarwal}\affiliation{Panjab University, Chandigarh 160014, India}
\author{Z.~Ahammed}\affiliation{Variable Energy Cyclotron Centre, Kolkata 700064, India}
\author{I.~Alekseev}\affiliation{Alikhanov Institute for Theoretical and Experimental Physics, Moscow, Russia}
\author{J.~Alford}\affiliation{Kent State University, Kent, Ohio 44242, USA}
\author{C.~D.~Anson}\affiliation{Ohio State University, Columbus, Ohio 43210, USA}
\author{A.~Aparin}\affiliation{Joint Institute for Nuclear Research, Dubna, 141 980, Russia}
\author{D.~Arkhipkin}\affiliation{Brookhaven National Laboratory, Upton, New York 11973, USA}
% ...

\collaboration{STAR Collaboration}\noaffiliation

\begin{abstract}
We report measurements of \upsi\ meson production 
in \pp, \dAu, and \AuAu\ collisions using the STAR detector at RHIC.
We compare the \upsi\ yield to the measured 
cross section in \pp\ collisions in order to quantify any modifications of the yield in cold nuclear matter using \dAu\ 
data and in hot nuclear matter using \AuAu\ data separated into three centrality classes. 
Our \pp\ measurement is based on three times the statistics of our previous result. 
We obtain a nuclear modification factor for \upsi(1S+2S+3S) in the rapidity range $|y|<1$ 
in \dAu\ collisions of \textcolor{black}{$\RdAu =  
0.79 \pm 0.24 (\mathrm{stat.}) \pm 0.03 (\mathrm{sys.}) \pm 0.10 (\pp\ \mathrm{sys.})$.}
A comparison with models including shadowing and initial state 
parton energy loss indicates the presence of additional cold-nuclear matter suppression.
Similarly, in the top 10\% most-central \AuAu\ collisions, we measure a nuclear modification factor of 
\textcolor{black}{$R_{AA}=0.49 \pm 0.1 \mathrm{(stat.)} \pm 0.02  \mathrm{(sys.)} \pm 0.06 (\pp\ \mathrm{ sys.)}$,}
which is a larger suppression factor than that seen in cold nuclear matter.
Our results are consistent with complete suppression of excited-state \upsi\ mesons in \AuAu\ collisions.
The additional suppression in \AuAu\ is consistent with the level 
expected in model calculations that include the presence of a hot, 
deconfined Quark-Gluon Plasma. 
However, understanding the suppression seen in \dAu\ 
is still needed before any definitive statements about the
nature of the suppression in \AuAu\ can be made.
\end{abstract}



\maketitle

\section{Introduction}\label{sec:Introduction}
In the study of the properties of the Quark-Gluon Plasma (QGP) 
an extensive effort has been devoted to
measuring quarkonium yields since these have been predicted to be sensitive to color deconfinement \cite{matsui}.
Studies have mainly focused on charmonium, but with the high collision energies
available at the Relativistic Heavy Ion Collider (RHIC) and the Large Hadron Collider (LHC)
 we can now study bottomonium in hot nuclear matter with sufficient statistics.
For a recent review of quarkonium in-medium,
see \eg\ Ref.~\cite{Brambilla:2010cs}, Sec.~5.  
One prediction is that excited quarkonium states are 
expected to dissociate at or above temperatures near that of the crossover to the \textcolor{black}{deconfined QGP phase, $\Tc \approx 
150-190\ \mev$} \cite{Karsch:2003jg,Cheng:2009zi,Borsanyi:2010bp}. The more tightly bound ground states are expected to dissociate at even 
higher temperatures. \textcolor{black}{The details of the temperature dependence of the dissociation of the excited states and of the feed-down pattern of the excited states into the ground state lead to a sequential 
suppression pattern of the inclusive upsilon states with increasing temperature \cite{Digal:2001ue}.}
The binding energy of the \upsitwo\ state ($\sim$540 MeV) is about half 
that of the \upsione\ state ($\sim$1.1 GeV); the \upsithree\ is still more weakly bound at $\sim$200 MeV.
Recent studies take into account not only the Debye screening effect on the heavy quark potential but also 
an imaginary part of the potential which modifies the widths of the various quarkonia states 
(\eg\ \cite{Laine:2006ns,Petreczky:2010tk,Dumitru:2009fy,Rothkopf:2011db}). 
In Ref.~\cite{Petreczky:2010tk} it is estimated that the \upsitwo\ state will 
melt at a temperature of $T\approx 250\ \mev$, whereas the ground state \upsione\ will melt at 
temperatures near $T\approx 450\ \mev$.  

We focus here on the measurement of bottomonium mesons in collisions at \sqrtsNN=200 \gev. 
An observation of suppression in the bottomonium sector in hot nuclear matter
is important for two reasons. First, it would be evidence for color 
deconfinement in the produced matter since the 
aforementioned effects are all ultimately based on studies of the high temperature 
phase of Quantum Chromodynamics (QCD) done on the lattice, where color
is an active degree of freedom. Second, bottomonium suppression provides a way to 
estimate model-dependent bounds on the temperature 
with the bounds depending on the particular suppression pattern seen.

The cross section for bottomonium 
production is smaller than that of charmonium \cite{Adare:2006kf,Adare:2009js,Abelev:2010am}
making the experimental 
study of \upsi\ production challenging. 
However, the theoretical interpretation of bottomonium 
suppression is less complicated than that of 
charmonium for several reasons.  
While charmonium production at RHIC and higher energies can be affected by the statistical recombination 
of charm quarks that are produced in 
different nucleon-nucleon collisions within the same nuclear interaction event, 
this effect is negligible for bottomonium due to the much smaller  
\bbbar\ production cross section 
($\sigma_{b\bar{b}}$ is measured to be in the range 1.34  -- 1.84 $\mu$b \cite{Agakishiev:2011mr} 
and calculated to be $1.87^{+0.99}_{-0.67}\ \mu$b \cite{Cacciari:2005rk}, 
compared to $\sigma_{c\bar{c}}\approx $ 550 -- 1400 $\mu$b 
\cite{Adams:2004fc,Adare:2010de}). Another 
complication in the charmonium case 
is that even in a purely hadronic scenario, 
charmonium mesons can be suppressed due to their interaction 
with hadronic co-movers \cite{Gerschel:1988wn,Gerschel:1998zi}. 
The cross section for inelastic collisions of \upsione\ with hadrons is small \cite{Lin:2000ke}.
Hence, absorption in the medium by the abundantly produced co-moving hadrons is
predicted to be minimal. 
The cold-nuclear-matter (CNM) effects on \upsi\ production, which are those
seen in $p$+A collisions and can be due to  
shadowing of the parton distribution functions in the nucleus or energy-loss in the nucleus, 
can still be important. There is evidence of some \upsi\
suppression in fixed target experiments at 800 GeV/c lab momentum from E772 \cite{Alde:1991sw}.  
However, the CNM suppression observed for \upsi\ is
smaller than that for \jpsi\ reported by NA50 \cite{Alessandro:2006jt}. 
For all these reasons, the \upsi\ family is expected to be a 
cleaner and more direct probe of
hot QCD, and of the corresponding color deconfinement effects.

In this letter we present measurements of \upsi\ production in \pp, \dAu, and \AuAu\ collisions
at \sqrtsNN=200 \gev\ via the \epluseminus\ 
decay channel obtained by the STAR 
experiment. We extract invariant cross sections for all three collision systems studied.
Using this \pp\ measurement as a baseline we obtain the nuclear 
modification factor (\RdAu\ and \RAA) of the three states combined: \upsi(1S+2S+3S).   
The ratio \RAA\ is used to 
quantify deviations of the yields in \dAu\ and \AuAu\ 
compared to those expected from a superposition of elementary \pp\ collisions. 
The data were taken during 2008 (\dAu), 2009 (\pp) and 2010 (\AuAu) at RHIC, 
and correspond to integrated luminosities of 28.2 nb$^{-1}$, 20.0 \invpb, and 1.08 nb$^{-1}$, respectively. 
All three datasets were taken with the same detector configuration.
For this reason the data from our previous p+p result (2006) was not
included in this analysis; the amount of material in the detector
at that point was substantially larger than it was in the three datasets
discussed here.
We compare our data to model calculations of the cross section based 
on perturbative QCD (pQCD)~\cite{Frawley:2008kk},
and to recent models of \upsi\ production
in \dAu\ and \AuAu\ collisions~\cite{Vogt:2012fba,Ferreiro:2011xy,Arleo:2012rs,Strickland:2011aa,Emerick:2011xu,Liu:2010ej}.

\section{Experimental Methods}
\label{sec:ExpMethods}
The main detectors used are the STAR Time Projection Chamber (TPC) \cite{tpc} for tracking and 
the STAR Barrel Electro-Magnetic Calorimeter (BEMC) \cite{emc} for triggering. Both the
TPC and BEMC are used for particle identification.  
The starting point is the STAR \upsi\ trigger whose main components are a 
fast hardware Level-0 (L0)  trigger, which fires when a 
tower in the BEMC has energy $E_{L0-BEMC}\ge 4.2$ GeV, and a software Level-2 (L2) trigger, 
which requires the presence of two high-energy clusters in the 
BEMC ($>$4.5 GeV and $>$3.0 GeV). The cluster pair must also
have an opening angle greater than $90^\circ$ and an invariant mass above 5 \gevcc\ (6.5 \gevcc) in \pp\ (\dAu).
Note that energy measured at the triggering level 
\textcolor{black}{is partially calibrated leading to small but random biases.} 
Hence, triggering thresholds are not precise in energy.
The \upsi\ trigger is required to be in coincidence 
with the STAR minimum bias trigger.  For \pp\ collisions 
the minimum bias trigger is based on the STAR Beam-Beam Counters, 
while for \dAu\ and \AuAu\ it is based on the STAR Zero-Degree Calorimeters (ZDC) 
and the Vertex-Position Detectors (VPD). 
The L0-L2 combination was used for the \dAu\ data in 2008 and for \pp\ data in 2009.
In the \AuAu\ 2010 run, an upgrade to the 
STAR data acquisition system allowed the processing of all the L0 triggers above 
the $E_{L0-BEMC}=4.2$ GeV threshold, thus removing the need for a Level-2 trigger.

Some of the key components common to all these analyses are the tracking, 
matching between TPC tracks and BEMC L0 and L2 clusters, 
and electron identification techniques. 
The main differences between the three datasets are summarized as follows.  
For \AuAu\ collisions we use the charged particle multiplicity 
measured in the TPC in order to determine the centrality of the collision. 
Using a Glauber model simulation, the multiplicity classes in the collision are used 
to estimate the average number of participants (\Npart) and number of binary collisions (\Ncoll). 
The trigger, tracking, and electron identification efficiencies in the \AuAu\ case were studied as a function of centrality \textcolor{black}{(see Tab.~\ref{tab:AuAuEffs}).}
The presence of the underlying \AuAu\ event background increases the energy measured in the calorimeter towers 
and results in a slight increase in the trigger efficiency with increasing \Npart\ (more central collisions).  Similarly, the 
increase in the track density in the TPC results in a decrease in the tracking efficiency which is especially noticeable at high \Npart.  
  We used the specific 
ionization of the tracks in the TPC gas (\dEdx) for electron identification. In addition, 
the projection of the track onto the location of the BEMC shower maximum position
was required to match the measured BEMC cluster position. 
Once a track was matched to a calorimeter cluster, 
the ratio of the  energy of the cluster to the TPC momentum ($E/p$) was also used for electron identification.
\textcolor{black}{The combined acceptance times efficiency for detecting an \upsi\ at mid-rapidity ($|y|<0.5$) in \AuAu\ taking into account all aforementioned effects was found to vary from $\sim12\%$ in peripheral collisions to $\sim9\%$ in central collisions.}

\begin{table}[t]
\begin{tabular}{l|c|c|c|}
Centrality & \Npart & Rapidity & Efficiency \\
\hline
\multirow{3}{*}{0-60\%} & \multirow{3}{*}{$162\pm9$} & $|y|<0.5$ & 0.122\\
\cline{3-4}
& & $0.5<|y|<1.0$ & 0.055\\
\cline{3-4}
& & $|y|<1.0$ & 0.088\\
\hline
\multirow{3}{*}{0-10\%} & \multirow{3}{*}{$326\pm4$} & $|y|<0.5$ & 0.089\\
\cline{3-4}
& & $0.5<|y|<1.0$ & 0.039\\
\cline{3-4}
& & $|y|<1.0$ & 0.064\\
\hline
\multirow{3}{*}{10-30\%} & \multirow{3}{*}{$203\pm10$} & $|y|<0.5$ & 0.125\\
\cline{3-4}
& & $0.5<|y|<1.0$ & 0.055\\
\cline{3-4}
& & $|y|<1.0$ & 0.089\\
\hline
\multirow{3}{*}{30-60\%} & \multirow{3}{*}{$80\pm10$} & $|y|<0.5$ & 0.126\\
\cline{3-4}
& & $0.5<|y|<1.0$ & 0.056\\
\cline{3-4}
& & $|y|<1.0$ & 0.091\\
\hline
\end{tabular}
\caption{\textcolor{black}{Upsilon reconstruction efficiency in \AuAu.
The total efficiency includes triggering efficiency, tracking efficiency, electron identification efficiency, and geometrical acceptance.}}
\label{tab:AuAuEffs}
\end{table}

The cuts used in these analyses were chosen such that the tracking and electron identification efficiencies would be
similar across the three datasets, allowing the systematic uncertainties to 
approximately cancel in the measurement of \RAA. 
For further detail, the techniques used in these \upsi\ measurements 
were described extensively for our previous \pp\ 
measurement~\cite{Abelev:2010am} based on a 7.9 \invpb\ dataset.
All evaluated sources of systematic uncertainty are summarized in Tab.~\ref{tab:SystUncertainty}.
An important effect in addition to those discussed in \cite{Abelev:2010am} is the change in tracking and mass resolution with increasing detector occupancy.
Simulated \upsi\ events were embedded in real data and their reconstructed line shapes were studied as a function of collision system and detector occupancy.
In \pp\ collisions, we find a mass resolution of 1.3\% for reconstructed \upsione.
Due to additional TPC
alignment errors for the \dAu\ and \AuAu\ data the mass resolutions
of the \upsione\ increased to 2.7\% in \dAu\ and peripheral \AuAu\ collisions
and 2.9\% in central collisions. This decreased mass resolution was
accounted for in the binary scaling estimates of \upsi(1S+2S+3S) yields
(see gray bands in Figs.~\ref{fig:InvariantMass1} and \ref{fig:InvariantMassAuAuCentBins}).
Systematic uncertainties in those scaling estimates (line shapes) are included in the errors in Table~\ref{tab:SystUncertainty}.
For all results we quote, the \upsi\ data are integrated over all transverse momenta. 

\begin{table}[t]
  \centering
  \begin{tabular}{l|c|}

    Source & Relative uncertainty \\ \hline
    Luminosity, Vernier scan (\pp)   & $\pm 7\%$ \\
    BBC efficiency                 & $\pm$ 9\%\\
    ZDC-Au trig. eff. (\dAu)  & $\pm 3\%$ \\
    Vertex finding eff. (\pp) 	& $\pm 1\%$ \\ 
    Vertex finding eff. (\dAu)   	& $\pm 0.1 \%$ \\
    Vertex finding eff. (\AuAu)   	& $\pm  0.1 \%$ \\
    \dAu\ min. bias $\sigma$	& $\pm4 \%$ \\
    Glauber model params.    & $+0.9\%, -0.5\%$ \\
    Acceptance  				& $+1.7\%$, $-3.0\%$\\
    L0 ADC threshold 		& $+8.7\%$, $-2.3\%$\\
    L2 $E_{\mathrm{cluster}}$	& $+1.2\%$, $-0.6$\% \\
    L2 $\cos\theta$ cut 		& $\sim 0\%$ \\
    L2 mass cut 				& $\sim 0\%$ \\
    Tracking efficiency 		& $\pm2\times5.88\%$\\
    Track-to-tower matching eff. & +0.2\%, -1.1\% \\
    $E/p$ cut efficiency 		& $\pm 3\%$ \\
    \dEdx\ cut efficiency & $\pm2\times 2.2\%$ \\
    \dAu\ excited state ratio	&   +0\%, -2\% \\
    \AuAu\ excited state ratio	&   $+1\%, -2\%$ \\
    \upsi\ \pp\ line shape       &  $+6.0\%$, $-4.1\%$  \\
    \upsi\ \dAu\ line shape     &  $+1.8\%$, $-1.2\%$  \\
    \upsi\ \AuAu\ line shape   &  $+0.8\%$, $-0.9\%$  \\
    \upsi\ \pT\ shape               & $\pm 1.7\%$ \\
    \upsione\ purity, \dAu\ and \AuAu          & +0\%, -7.5\% \\
    \hline
    Total syst., $\sigma_{pp}$			&  $+21.1\%$,\  $-19.0\%$ \\
    Total syst., $\sigma_{d\mathrm{Au}}$ 			&  $+17.5$\%,\ $-15.6\%$ \\
    Total syst., $\sigma_{\mathrm{AuAu}}$ 			&  $+16.0$\%,\ $-14.1\%$ \\
    \hline
    Common normalization syst.					& $+12.9\%,\ -12.2\%$ \\
    \RdAu, syst. 				&    $+3.5\%$,\ $-3.8\%$\\
    \RdAu(1S), syst. 				&    $+3.5\%$,\ $-8.4\%$\\
    \RAA, syst.				&    $+3.2\%$,\ $-3.6\%$\\
    \RAA(1S), syst.				&    $+3.2\%$,\ $-8.3\%$\\
    \hline
    \end{tabular}
  \caption{Systematic uncertainties affecting the cross sections and ratios.
           \textcolor{black}{Uncertainties stemming from TPC momentum resolution are included in the line shape uncertainties.}}
  \label{tab:SystUncertainty}
\end{table}

\section{Results and Discussion}
\begin{figure}[t!]
\begin{center}
\includegraphics[width=0.5\textwidth]{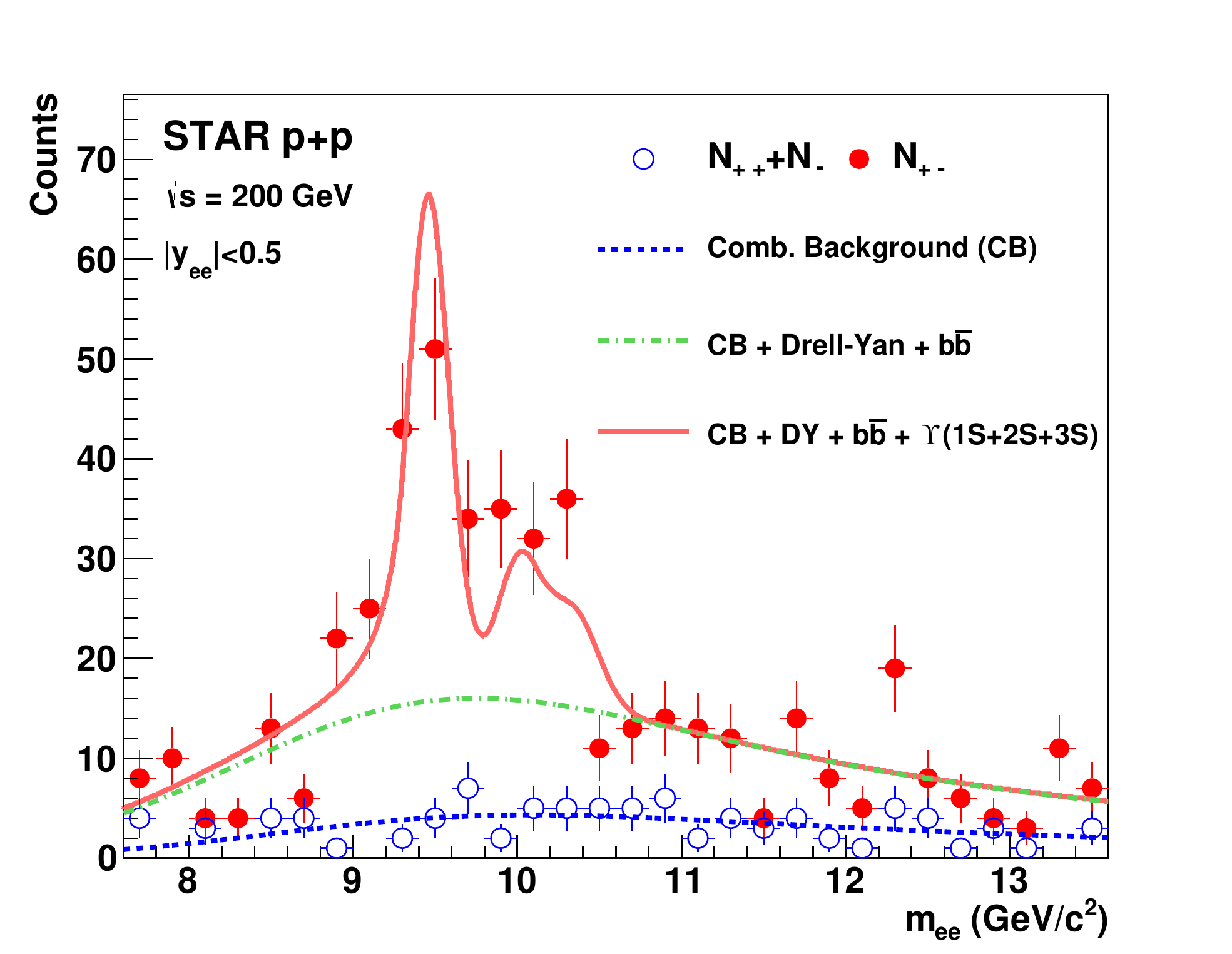}
\includegraphics[width=0.5\textwidth]{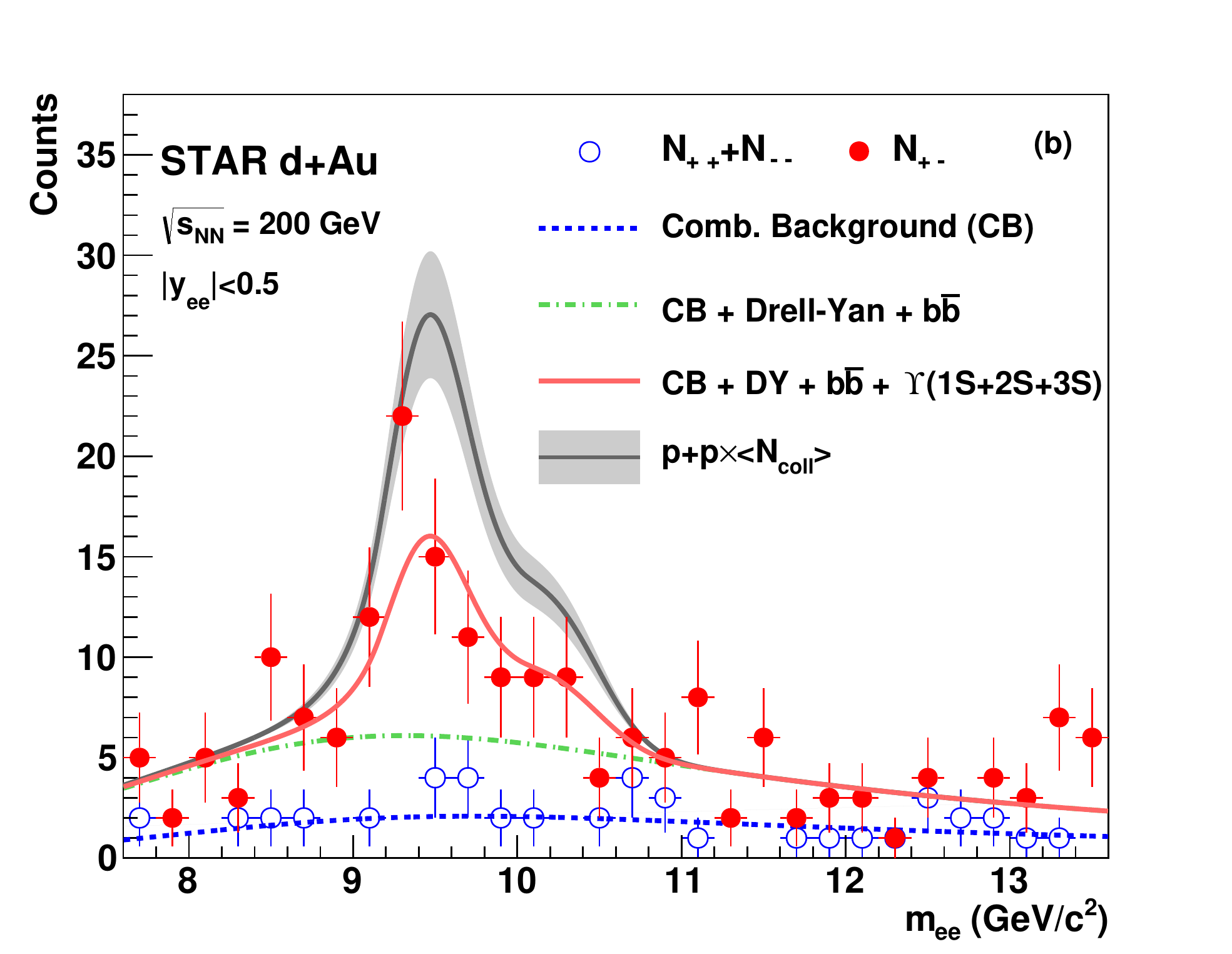}
\caption{(Color online) Invariant mass distributions of electron pairs in the region $|y_{\text{ee}}|<0.5$. (a): \pp. (b): \dAu.
Unlike-sign pairs are shown as red filled circles and like-sign pairs as hollow blue circles. The gray band shows the expected yield if \RdAu=1 \textcolor{black}{including resolution effects}. See text for description of yield extraction.}
\label{fig:InvariantMass1}
\end{center}
\end{figure}

Figure~\ref{fig:InvariantMass1} shows the invariant mass distributions 
of electron pairs for \pp\ (top) and \dAu\ (bottom) in the kinematic region $|y_{\Upsilon}|<0.5$.
\label{sec:Results}
Unlike-sign pairs are shown as red filled circles and like-sign pairs as hollow blue circles. 

The data are fit with a parameterization consisting of the sum of various contributions to the
electron-pair invariant-mass spectrum.
\textcolor{black}{The fit is performed simultaneously with the like-sign and unlike-sign spectra using a maximum-likelihood method.}
The lines in Fig.~\ref{fig:InvariantMass1} show the yield from the combinatorial background
(dashed blue line), the result of adding 
the physics background from Drell-Yan and \bbbar\ pairs
(dot-dashed green line), and finally the inclusion of the \upsi\ contribution 
(solid red line). 
The shape of the Drell-Yan continuum is obtained via a 
next-to-leading order (NLO) pQCD calculation from Vogt \cite{Vogt:DY}. PYTHIA 8 was 
used to calculate the shape of the \bbbar\ contribution \cite{Sjostrand}.
We model each of the $\Upsilon$ states with a Crystal Ball function 
\cite{CBFunction}, which incorporates detector resolution and 
losses from bremsstrahlung in the detector material.

The fit is done to the unlike and like-sign data simultaneously. The fit to the combinatorial 
background component extracted from the like-sign data is shared by the functional form 
used to parameterize the unlike-sign data.
In the usual like-sign subtraction procedure some information would be lost.
In contrast, by performing a simultaneous fit to 
both the like-sign and unlike-sign signals we optimize the statistical power of our data.
The L2 trigger condition has the effect of cutting off the lower 
invariant masses. This cut-off shape is parameterized 
in the fits using an error function. 

We integrate the unlike-sign invariant mass distribution in 
the region $8.8 - 11\ \gevcc$ and subtract from the data the fit to the 
combinatorial, Drell-Yan, and \bbbar\ background components 
in order to obtain the yield of \upsi(1S+2S+3S). 
After accounting for efficiencies and sampled luminosity, 
we calculate a production cross section in \pp\ collisions of: 
\textcolor{black}{$B_{ee}\times d\sigma/dy|_{|y|<0.5}=64\pm10 (\mathrm{stat.+fit})^{+14}_{-12}$ pb. 
Our previous result of $114\pm38^{+23}_{-24}$ pb \cite{Abelev:2010am}
is consistent with our new measurement.
The greater sampled luminosity and decreased detector material in 2009 led to improved statistics and lower systematic uncertainties in the present measurement.}

\textcolor{black}{In Fig.~\ref{fig:InvariantMass1}(b), the gray band 
shows the expected signal from the \pp\ data scaled by the number of binary collisions.  
Due to differences in detector occupancy and detector calibrations 
the width of the \upsi\ signal differs between collision systems and centralities. 
\textcolor{black}{As discussed in the previous section, a \textcolor{black}{misalignment in the TPC} in the \dAu\ and \AuAu\ datasets led to a broadening of the \upsi\ line shapes compared to the \pp\ dataset.}
This can be seen by examining the line shapes for the \upsi\ states in Fig.~\ref{fig:InvariantMass1}(a) (\pp) and Fig.~\ref{fig:InvariantMass1}(b) (\dAu).
\textcolor{black}{The average detector occupancy is comparable between the two systems, however the \dAu\ dataset has a noticeably broader line shape due to the aforementioned differences in calibration.
The effects of the broadening of the line shapes are taken into account in systematic uncertainties (Tab.~\ref{tab:SystUncertainty}).}
The comparison of the gray band \textcolor{black}{with} the \dAu\ data in panel (b) 
indicates a suppression of \upsi\ production with respect 
to binary-collision scaling.}

A similar procedure is followed for the region $0.5<|y_{\Upsilon}|<1$ in \pp\ collisions. 
We combine the results to obtain the differential cross section: \textcolor{black}{$B_{ee}\times d\sigma/dy |_{|y|<1}= 58 \pm 12 (\mathrm{stat.+fit}) ^{+12}_{-11}$ pb.}
In \dAu\ collisions, 
we analyze the yields separately in the regions $-1 < y_{\Upsilon} < -0.5$ and $0.5 < y_{\Upsilon} < 1$ 
because the \dAu\ system is not symmetric about $y=0$.  
Hence, averaging between forward and reverse rapidities is not warranted as it is in \pp.
Throughout this paper, the positive rapidity region is the deuteron-going direction, and the
negative rapidity region is the Au-going direction.
Integrating over our measured range ($|y_{\Upsilon}|<1$), the cross section in \dAu\ collisions is found to be
\textcolor{black}{$B_{ee}\times d\sigma/dy |_{|y|<1}= 19 \pm 3 (\mathrm{stat.+fit}) \pm 3  (\mathrm{syst}.)$ nb.}

We extract the \upsione\ yield directly by integrating over a narrower
mass window (8.8-9.8 GeV/c$^2$).
This mass window was chosen due to its high acceptance rate for \upsione\ and its high
rejection rate for the excited states. To account for sensitivity to the shape of the
\upsi\ signal, we varied the parameters of the line shape \textcolor{black}{obtained from simulations and data-driven methods discussed previously by their measured uncertainties} \textcolor{black}{and varied the excited states from unsuppressed to completely suppressed.}
We then recalculated both efficiency and purity (see Tab.~\ref{tab:SystUncertainty}, \upsione\ purity).
Those variations were taken into account as additional systematics when quoting \upsione\ results.

\begin{figure}[b!]
\includegraphics[width=0.47\textwidth, trim=0cm 0cm 0cm 0cm]{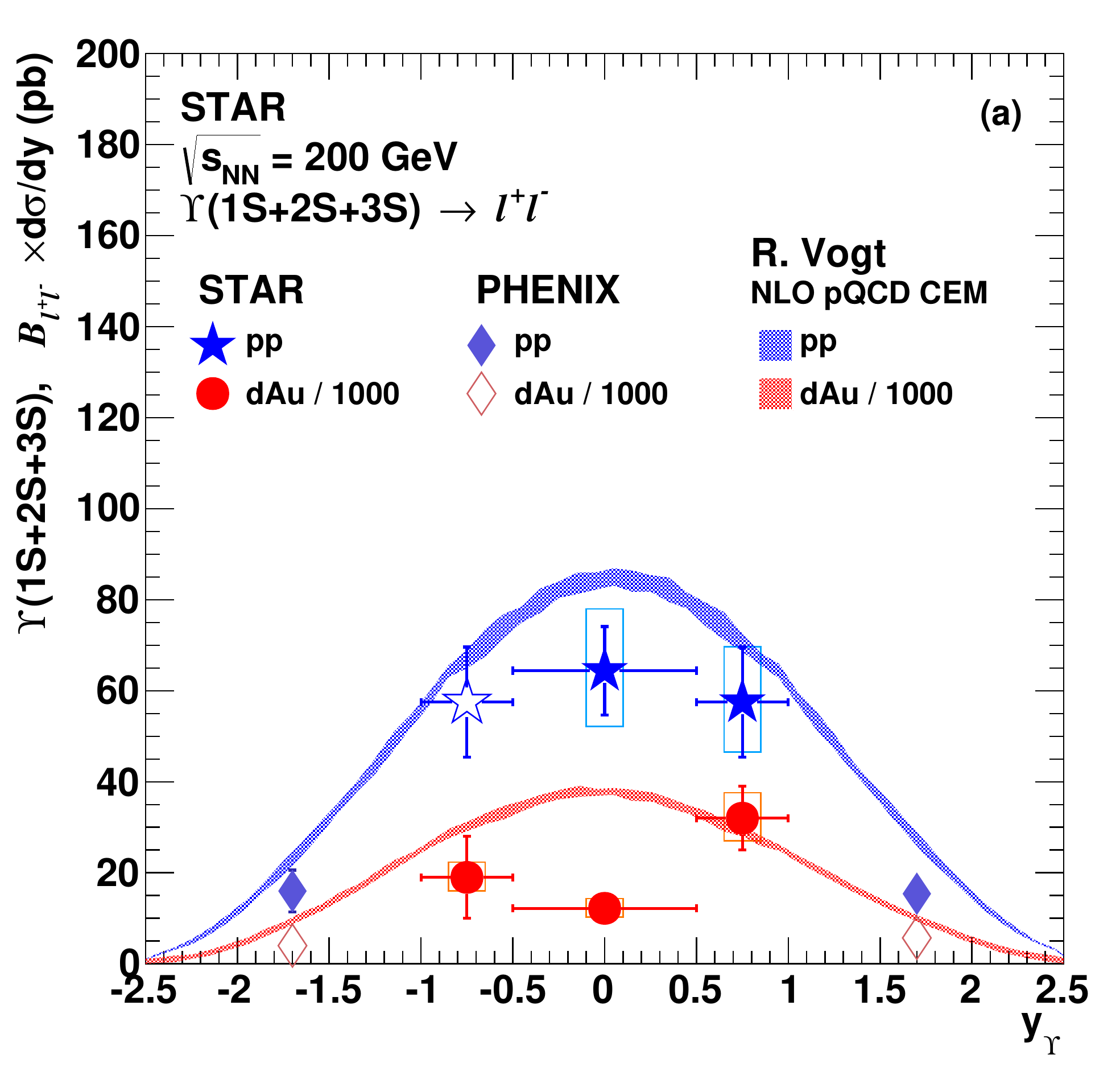}
\includegraphics[width=0.462\textwidth, trim=-0.5cm 0cm 0cm 0cm]{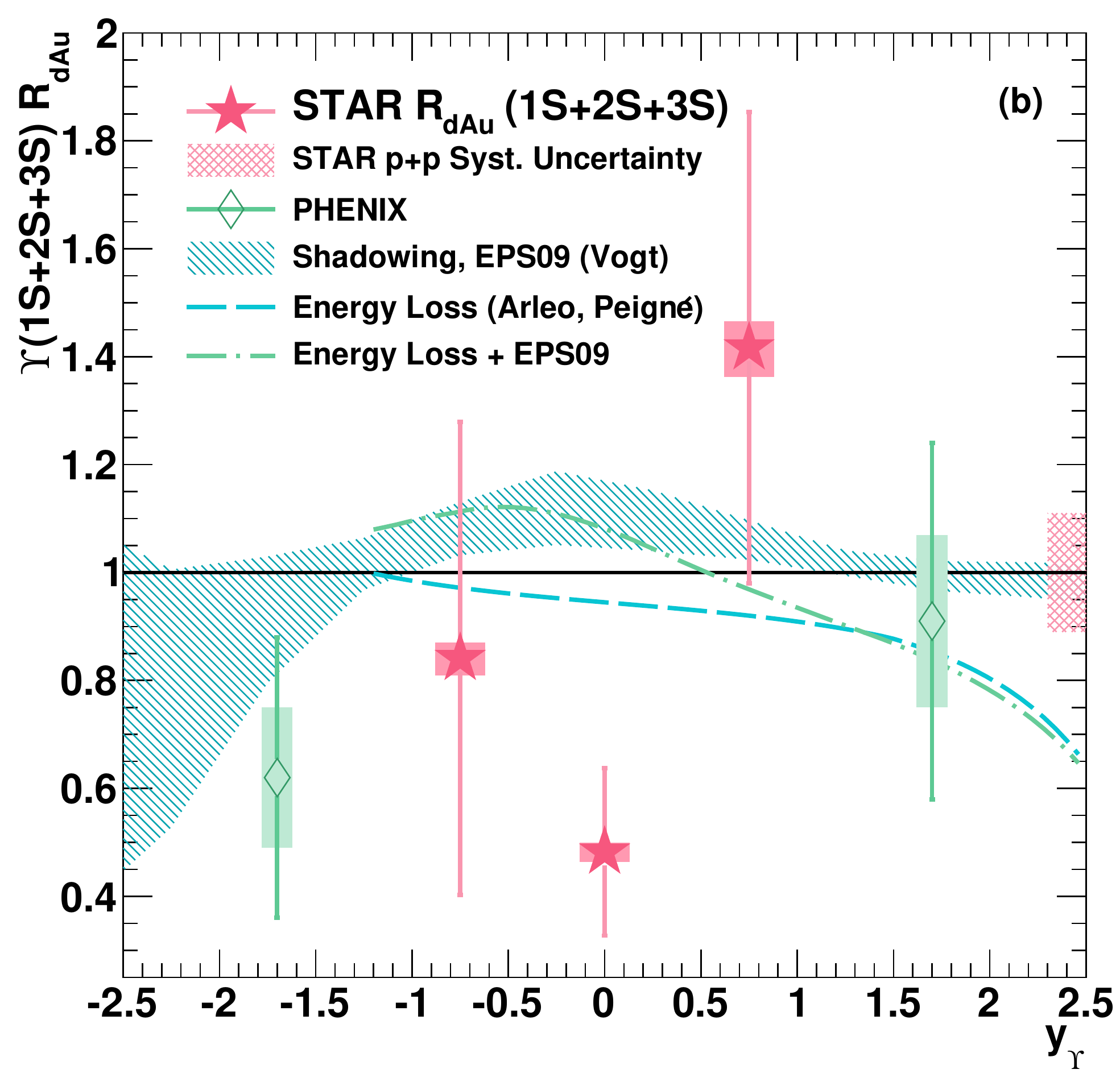}
\caption{(Color online) (a) $B_{ee}\times d\sigma/dy$ vs. $y$ for \pp\ collisions (blue stars) and for \dAu\ collisions (red, filled circles; scaled down by $10^3$).
Note that the hollow star at $y=-0.75$ is a reflection of the filled one at $y=0.75$ since 
these are not independent measurements. Results obtained by PHENIX are shown as filled diamonds. 
\textcolor{black}{Systematic errors are shown as boxes around the data.}
The shaded bands are from next-to-leading-order pQCD color evaporation model calculations. The \dAu\ 
prediction uses the EPS09 nPDF which includes shadowing \cite{Vogt:2012fba}. (b) \RdAu\ vs. $y$ for STAR (red stars) and PHENIX (green diamonds) results. \textcolor{black}{The band on the right shows the overall normalization uncertainty for the STAR results due to systematic uncertainties in the \pp\ measurement.} The shaded band shows the prediction for \RdAu\ from EPS09 and its uncertainty.
\textcolor{black}{The dashed curve shows suppression due to initial-state parton energy loss and the dot-dashed curve shows the same model with EPS09 incorporated \cite{Arleo:2012rs}.}}
\label{fig:UpsilonRapidity}
\end{figure}

Figure~\ref{fig:UpsilonRapidity}(a) shows the extracted \upsi(1S+2S+3S) 
cross section for \pp\ and \dAu\ as a function of rapidity. 
The \pp\ measurements are shown as blue stars and the \dAu\ measurements as red circles.
The \pp\ result in the region $0.5<|y_{\Upsilon}|<1.0$ is displayed as a star at $y=0.75$ and also as a hollow star at $y=-0.75$ to illustrate that the latter is not an independent measurement.
The data from PHENIX at forward rapidity for \pp\ (filled blue diamonds) and
 \dAu\ (hollow red diamonds) are also
shown \cite{Adare:2012bv}.

The cross sections in \pp\ are compared to an NLO pQCD calculation
of \upsi\ production in the Color Evaporation Model (CEM) \cite{Frawley:2008kk}, 
which is consistent with our data within the statistical and systematic uncertainties.
\textcolor{black}{The same calculation is performed for \dAu\ including shadowing effects \cite{Vogt:2012fba}.
The EPS09 set of nuclear Parton Distribution Functions (nPDF)  \cite{Eskola:2009uj} were used}.  
The model is in agreement with our data except for the mid-rapidity point which is lower
than the prediction.  To study this observation for \dAu\ further, we make a closer comparison to 
models and to previous measurements of \upsi\ production in p+A collisions. 

\begin{table*}[t!]
  \begin{tabular}{r | r | r | c | c}
     System & Centrality & States & Rapidity & $R_{AA,dA}$\\ 
     \hline 
      \multirow{8}{*}{\dAu} & \multirow{8}{*}{Min. bias} & \multirow{4}{*}{1S+2S+3S} & -1.0$<y_{\varUpsilon}<$-0.5 & 0.84 $\pm$ 0.40  $\pm$ 0.18  $\pm$ 0.03  $\pm$ 0.10  \\ 
      & & & $|y_{\varUpsilon}|<$0.5 & 0.48 $\pm$ 0.14  $\pm$ 0.07  $\pm$ 0.02  $\pm$ 0.06  \\ 
      & & & $0.5<y_{\varUpsilon}<1.0$ & 1.42 $\pm$ 0.32  $\pm$ 0.30  $\pm$ 0.05  $\pm$ 0.17  \\ 
      & & & $|y_{\varUpsilon}|<$1.0 & 0.79 $\pm$ 0.14  $\pm$ 0.10  $\pm$ 0.03  $\pm$ 0.09  \\ 
      \cline{3-5} 
      & & \multirow{4}{*}{1S}& -1.0$<y_{\varUpsilon}<$-0.5 & 0.74 $\pm$ 0.34  $\pm 0.16 ^{+0.03}_{-0.06}  \pm$ 0.09  \\ 
      & & & $|y_{\varUpsilon}|<$0.5 & 0.63 $\pm$ 0.18  $\pm 0.09 ^{+0.02}_{-0.05}  \pm$ 0.08  \\ 
      & & & 0.5$<y_{\varUpsilon}<$1.0 & 1.31 $\pm$ 0.29  $\pm 0.28 ^{+0.05}_{-0.11}  \pm$ 0.16  \\ 
      & & & $|y_{\varUpsilon}|<$1.0 & 0.83 $\pm$ 0.20  $\pm 0.11 ^{+0.03}_{-0.07}  \pm$ 0.10  \\ 
      \hline 
      \multirow{16}{*}{\AuAu} & \multirow{4}{*}{0-10\%} & \multirow{2}{*}{1S+2S+3S} & $|y_{\varUpsilon}|<$0.5 & 0.46 $\pm$ 0.05  $\pm$ 0.07  $\pm$ 0.02  $\pm$ 0.05  \\ 
      & & & $|y_{\varUpsilon}|<$1.0 & 0.49 $\pm$ 0.13  $\pm$ 0.07  $\pm$ 0.02  $\pm$ 0.06  \\ 
      \cline{3-5} 
      & & \multirow{2}{*}{1S} & $|y_{\varUpsilon}|<$0.5 & 0.69 $\pm$ 0.05  $\pm 0.10 ^{+0.02}_{-0.06}  \pm$ 0.08  \\ 
      & & & $|y_{\varUpsilon}|<$1.0 & 0.66 $\pm$ 0.13  $\pm 0.10 ^{+0.02}_{-0.05}  \pm$ 0.08  \\ 
      \cline{2-5} 
      \ & \multirow{4}{*}{10-30\%} & \multirow{2}{*}{1S+2S+3S} & $|y_{\varUpsilon}|<$0.5 & 0.69 $\pm$ 0.16  $\pm$ 0.10  $\pm$ 0.02  $\pm$ 0.08  \\ 
      & & & $|y_{\varUpsilon}|<$1.0 & 0.82 $\pm$ 0.20  $\pm$ 0.12  $\pm$ 0.03  $\pm$ 0.10  \\ 
      \cline{3-5} 
      & & \multirow{2}{*}{1S} & $|y_{\varUpsilon}|<$0.5 & 0.85 $\pm$ 0.16  $\pm 0.13 ^{+0.03}_{-0.07}  \pm$ 0.10  \\ 
      & & & $|y_{\varUpsilon}|<$1.0 & 1.07 $\pm$ 0.20  $\pm 0.16 ^{+0.03}_{-0.09}  \pm$ 0.13  \\ 
      \cline{2-5} 
      \ & \multirow{4}{*}{30-60\%} & \multirow{2}{*}{1S+2S+3S} & $|y_{\varUpsilon}|<$0.5 & 0.74 $\pm$ 0.22  $\pm$ 0.11  $\pm$ 0.03  $\pm$ 0.09  \\ 
      & & & $|y_{\varUpsilon}|<$1.0 & 0.82 $\pm$ 0.22  $\pm$ 0.12  $\pm$ 0.03  $\pm$ 0.10  \\ 
      \cline{3-5} 
      & & \multirow{2}{*}{1S} & $|y_{\varUpsilon}|<$0.5 & 1.22 $\pm$ 0.22  $\pm 0.18 ^{+0.04}_{-0.10}  \pm$ 0.15  \\ 
      & & & $|y_{\varUpsilon}|<$1.0 & 1.19 $\pm$ 0.22  $\pm 0.18 ^{+0.04}_{-0.10}  \pm$ 0.14  \\ 
      \cline{2-5} 
      \ & \multirow{4}{*}{0-60\%} & \multirow{2}{*}{1S+2S+3S} & $|y_{\varUpsilon}|<$0.5 & 0.62 $\pm$ 0.11  $\pm$ 0.09  $\pm$ 0.02  $\pm$ 0.07  \\ 
      & & & $|y_{\varUpsilon}|<$1.0 & 0.66 $\pm$ 0.09  $\pm$ 0.10  $\pm$ 0.02  $\pm$ 0.08  \\ 
      \cline{3-5} 
      & & \multirow{2}{*}{1S} & $|y_{\varUpsilon}|<$0.5 & 0.85 $\pm$ 0.11  $\pm 0.13 ^{+0.03}_{-0.07}  \pm$ 0.10  \\ 
      & & & $|y_{\varUpsilon}|<$1.0 & 0.88 $\pm$ 0.09  $\pm 0.13 ^{+0.03}_{-0.07}  \pm$ 0.11  \\ 
      \hline 
  \end{tabular}
  \caption{Table of \RdAu\ and \RAA\ results. The results are listed in the form $a \pm b \pm c \pm d \pm e$ where $a$ is \RdAu\ or \RAA, $b$ is the \dAu\ or \AuAu\ statistical uncertainty, $c$ is the \pp\ statistical uncertainty, $d$ is the \dAu\ or \AuAu\ systematic uncertainty, and $e$ is the \pp\ systematic uncertainty.}
  \label{tab:RAA}
\end{table*}

To focus on expected shadowing effects, we obtain
the nuclear modification factor \RdAu\ as a function of rapidity.  The nuclear modification factor is defined in nucleus-nucleus collisions as
\[
    R_{AA} = \frac{1}{\frac{\sigma_{AA}}{\sigma_{pp}} } \times \frac{1}{<N_{coll}>} \times \frac{B_{ee}\times\left(\frac{d\sigma_{AA}}{dy}\right)^{\Upsilon} }{B_{ee}\times\left(\frac{d\sigma_{pp}}{dy}\right)^{\Upsilon} }
\]
where the first factor accounts for the difference in inelastic cross section in
\pp\ to \dAu\ or \AuAu\ collisions.
The second factor accounts for the average number of nucleon collisions in a \dAu\ or \AuAu\ collision as calculated by a Glauber model.
The third factor accounts for the measured \upsi\ production in \pp, \dAu\ or \AuAu.
\textcolor{black}{We used the following total inelastic cross sections: $\sigma_{pp}=42$ mb, $\sigma_{d\textrm{Au}}=2.2$ b, and $\sigma_{\textrm{AuAu}}=6$ b.}

Our results for \RdAu\ are shown in Fig.~\ref{fig:UpsilonRapidity}(b) and summarized in Tab.~\ref{tab:RAA}. Our data (red stars) 
are compared to
CEM calculations with the uncertainty from the EPS09 nPDF shown as the shaded
region.
Note that this prediction for \RdAu, which includes modification of the nuclear PDFs but does not include absorption, 
implies a modification factor of $\RdAu\approx1.1$.
A calculation in Ref.~\cite{Ferreiro:2011xy} explored various nPDFs (EKS98, EPS08, and nDSg)
and also gave \RdAu\ values
above 1 with enhancements in the range of 5-20\%.  The models
are in agreement with the data except in the $y\sim0$ region. An additional effect which can
suppress the \upsi\ yield is initial-state parton energy loss. A calculation by 
Arleo and Peign\'{e} \cite{Arleo:2012rs} incorporating this effect is shown
as the dashed line.  The calculation for a combination of energy loss 
and shadowing using EPS09 is shown as the dashed-dotted line.
The energy-loss model is also in agreement with the data except for the mid-rapidity point. 
The model from \cite{Arleo:2012rs} does not include absorption from interactions with spectator nucleons.
However, those effects only play a role in the rapidity region $y\lesssim 1.2$, where the \upsi\ mesons
would be closer to the frame of the Au spectators. Therefore, the suppression at mid-rapidity is indicative of effects
beyond shadowing, initial-state parton energy loss, or absorption by spectator nucleons.

\begin{figure}[ht!]
\includegraphics[trim=1.2cm 0cm -0.6cm 0cm, width=0.43\textwidth]{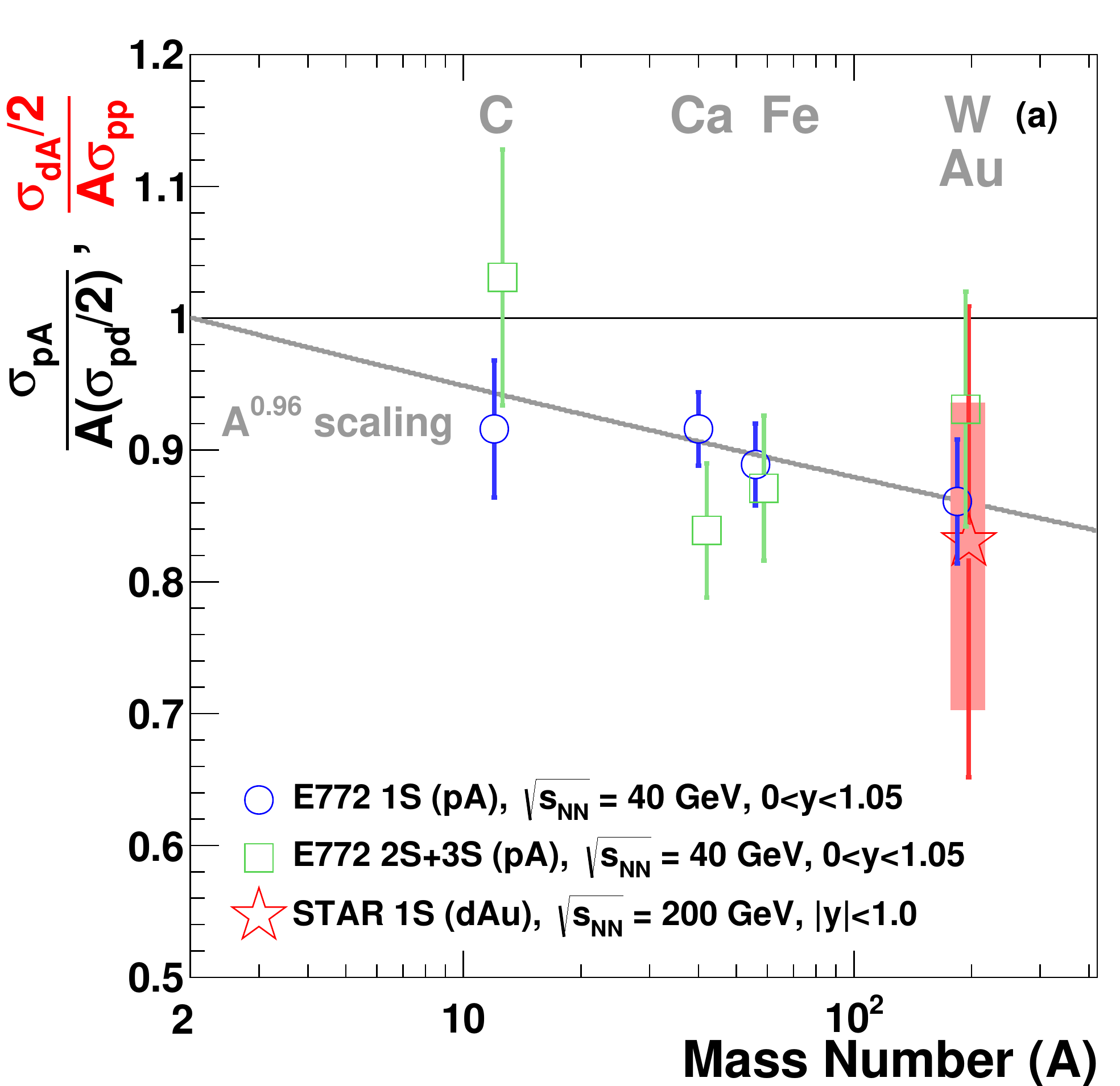}
\includegraphics[trim=-0.3cm 0cm 0cm 0cm, width=0.43\textwidth]{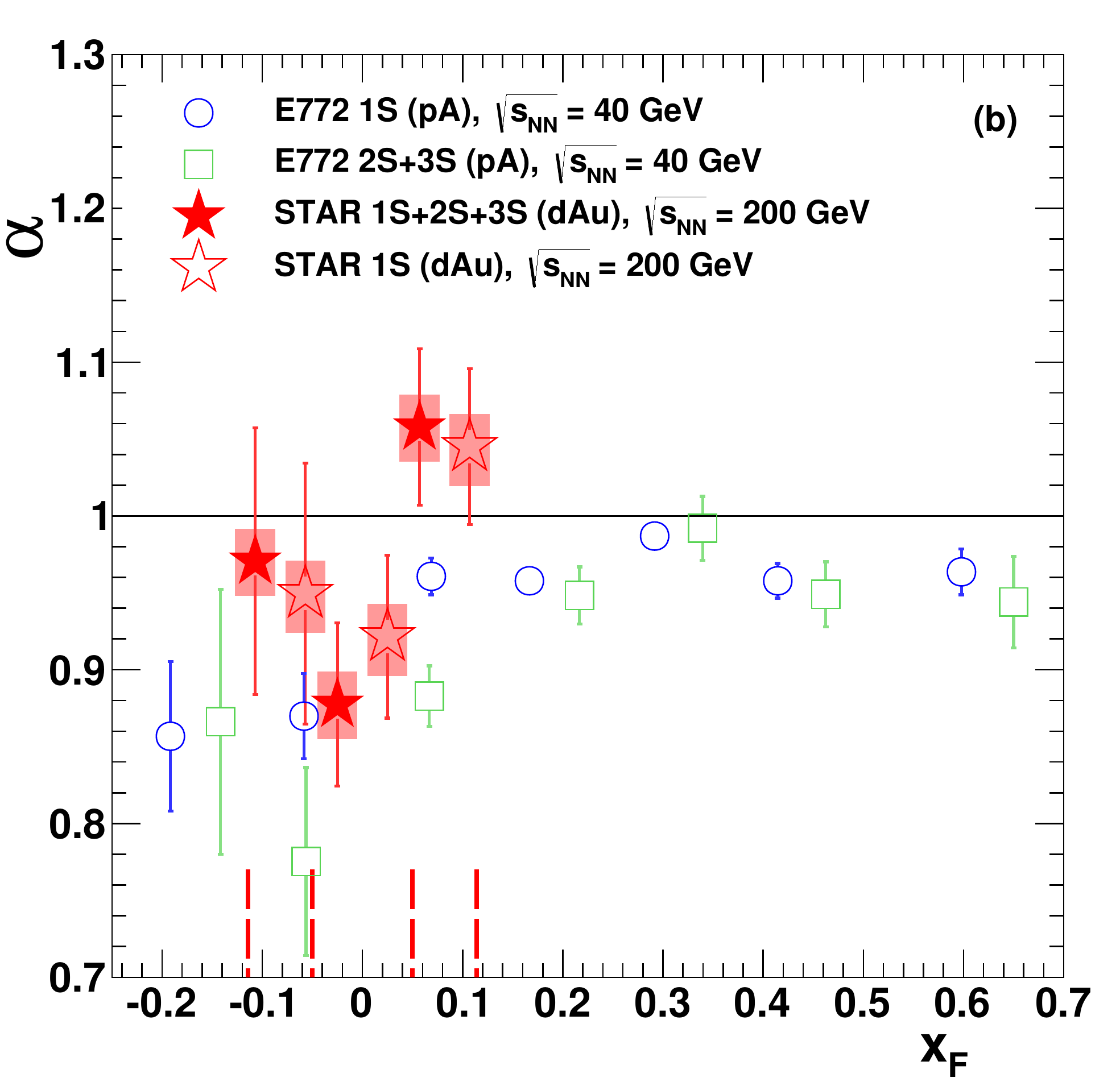} 
\caption{(Color online) Comparison of our \dAu\ measurements to the pA measurements from E772. 
(a): Ratio of $\Upsilon$ production in pA to pp scaled by mass number as a function of mass number. 
Shown are the 1S (hollow blue circles) and 2S+3S (hollow green squares) $\Upsilon$ 
measurements from E772 and our \textcolor{black}{1S} measurement (red star).  
Also shown is the model used by E772 where $\sigma_{pA}=A^{\alpha}\sigma_{pp}$. 
E772 found $\alpha=0.962\pm0.006$\cite{Alde:1991sw}. (b): Exponent $\alpha$ as a function of $x_{F}$. \textcolor{black}{The vertical, dashed red lines at the bottom of the plot denote the width of the $x_{F}$ bins for the STAR measurements. Note that the STAR data points are offset within the bins for clarity.}}
\label{fig:E772}
\end{figure}

\textcolor{black}{We compare our measurements with results from E772 at $\sqrtsNN=40 $ GeV,
where suppression of the \upsi\ states in $\mbox{$p$+$A$}$ was observed. 
This is illustrated in Fig.~\ref{fig:E772}(a), which shows the ratio of the cross section in \dAu\ collisions for STAR ($p$+$A$ for E772)
to that of \pp\ collisions normalized by the mass number $A$.  E772 plotted a ratio of 
extracted cross sections normalized by the data where the proton beam hit a liquid 
deuterium target ($A=2$). Assuming that the cross section scales as
$\sigma_{pA} = A^{\alpha}\sigma_{pp}$, and using their $p$+$d$ result as the baseline, the
solid line shows that the ratio should scale as $(A/2)^{\alpha-1}$.  Our measurement in \dAu\ for 
the \upsione\ state (red star) is consistent with the fit to the 
E772 data, shown as hollow blue circles for \upsione\ and hollow green squares for \upsi(2S+3S).
Our results cover the rapidity range $|y|<1$ whereas the E772 measurements were 
in the forward region $0 < y < 1.05$.
To better compare our rapidity coverage, we plot the $\alpha$ value as a function of Feynman-x ($x_F$) in 
Fig.~\ref{fig:E772}(b).}
The larger suppression we observe at mid-rapidity
is also consistent with the larger suppression seen in E772 for $x_F\sim0$.

\begin{figure}[h!]
\begin{center}
\includegraphics[width=0.465\textwidth]{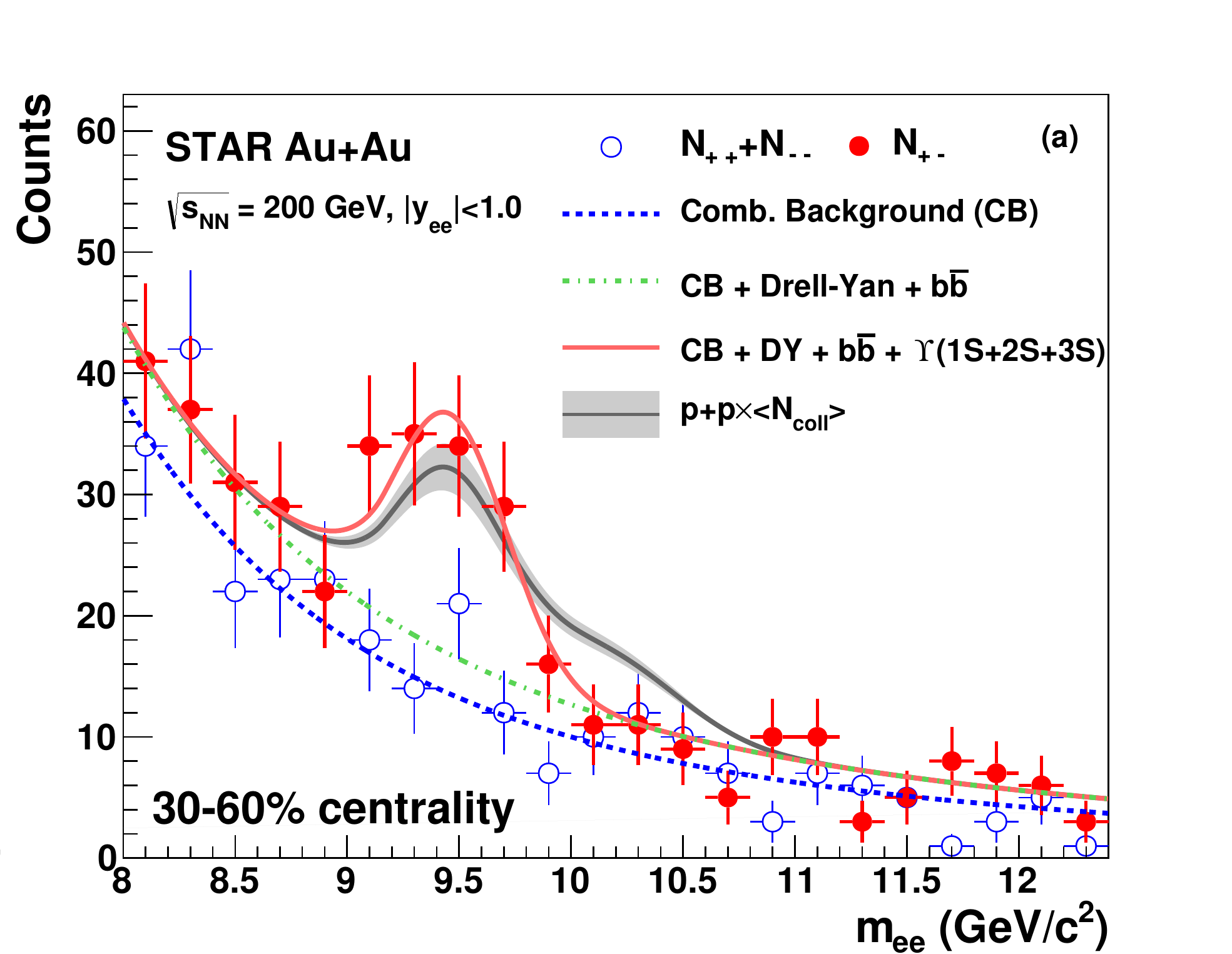}
\includegraphics[width=0.465\textwidth]{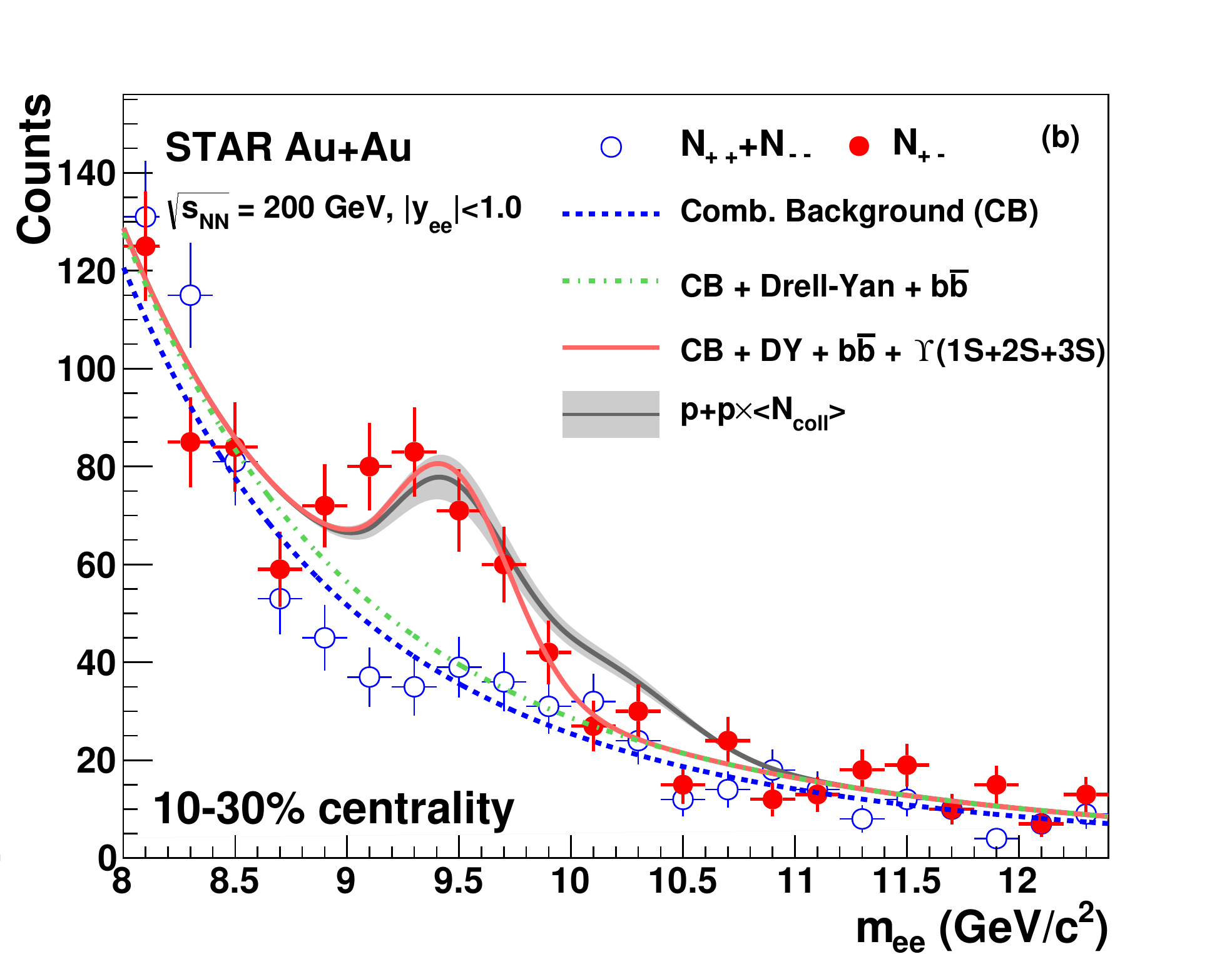}
\includegraphics[width=0.465\textwidth]{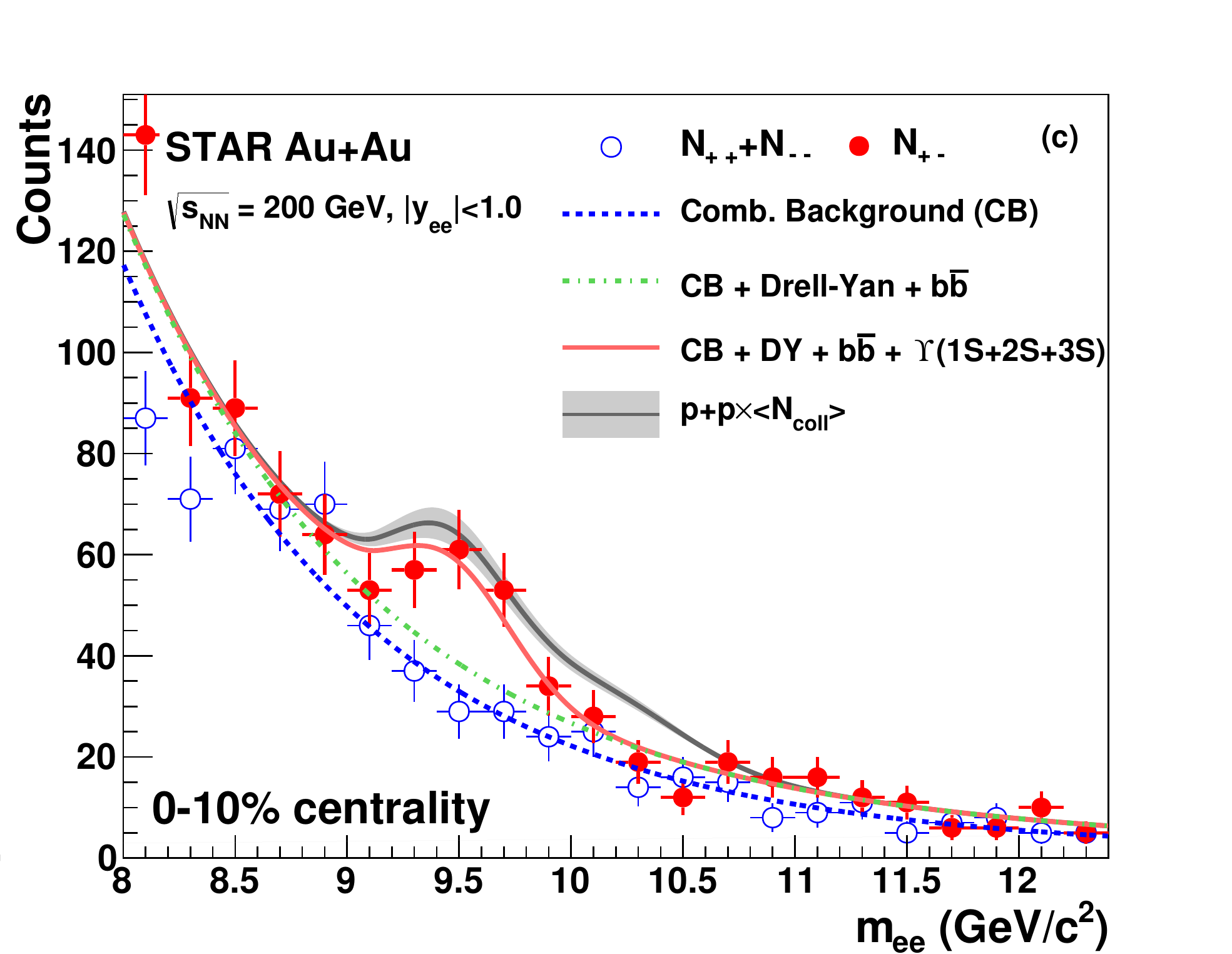}
\caption{(Color online) Invariant mass distributions of electron pairs in the region $|y_{\text{ee}}|<1.0$ 
for the centrality selections 30-60\% (a), 10-30\% (b), 
and 0-10\% (c). 
Unlike-sign pairs are shown as filled red circles and like-sign pairs as hollow blue circles. 
Fits are described in the text. 
The gray band shows the expected signal assuming scaling of the \pp\  yield with the number of binary collisions \textcolor{black}{including resolution effects.}}
\label{fig:InvariantMassAuAuCentBins}
\end{center}
\end{figure}

We next turn to the measurements in \AuAu\ collisions. 
The \AuAu\ invariant mass spectrum is fit in 3 centrality bins: 
30-60\% (Fig. \ref{fig:InvariantMassAuAuCentBins}(a), 
10-30\% (Fig. \ref{fig:InvariantMassAuAuCentBins}(b), 
and 0-10\% (Fig. \ref{fig:InvariantMassAuAuCentBins}(c).  
As in Fig.~\ref{fig:InvariantMass1} we show the fits including, in succession, combinatorial 
background (dashed blue line), the physics background from Drell-Yan and \bbbar\ pairs (dot-dashed green line), and the \upsi\ contribution 
(solid red line). The absence of the L2 trigger in the 
\AuAu\ dataset removes the cut-off effect. One can therefore see the background \textcolor{black}{(modeled as the sum of two exponentials)}, 
dominated by the combinatorial component, 
rising at lower invariant  mass. Measured cross sections are summarized in Tab.~\ref{tab:crossSec}.
The gray bands in the \AuAu\ figure illustrate the expected signal from the \pp\ data 
scaled by the number of binary collisions. There is a clear 
suppression of the expected yield in \AuAu\ collisions.

\begin{table}[!h]
  \begin{tabular}{r | c | c}
     Centrality & Rapidity & $d\sigma/dy$ (nb) \\ 
     \hline 
      \multirow{2}{*}{0-60\%} & $|y_{\varUpsilon}|<0.5$ & 2170 $\pm$ 357 $\pm$ 349  \\ 
      & $|y_{\varUpsilon}|<1.0$ & 2180 $\pm$ 250 $\pm$ 351  \\ 
      \hline 

      \multirow{2}{*}{0-10\%} & $|y_{\varUpsilon}|<0.5$ & 3950 $\pm$ 416 $\pm$ 636  \\ 
      & $|y_{\varUpsilon}|<1.0$ & 3990 $\pm$ 1020 $\pm$ 642  \\ 
      \hline 

      \multirow{2}{*}{10-30\%} & $|y_{\varUpsilon}|<0.5$ & 3040 $\pm$ 676 $\pm$ 489  \\ 
      & $|y_{\varUpsilon}|<1.0$ & 3430 $\pm$ 827 $\pm$ 552  \\ 
      \hline 

      \multirow{2}{*}{30-60\%} & $|y_{\varUpsilon}|<0.5$ & 905 $\pm$ 225 $\pm$ 146  \\ 
      & $|y_{\varUpsilon}|<1.0$ & 950 $\pm$ 198 $\pm$ 153  \\ 
      \hline 

  \end{tabular}
  \caption{\upsi\ production cross sections in \AuAu\ collisions. The first uncertainty listed is the combination of the statistical and fit uncertainties and the second is the systematic uncertainty.}
  \label{tab:crossSec}
\end{table}

\begin{figure}[hb]
\includegraphics[width=0.40\textwidth, trim=1.8cm 0cm 0cm 0cm]{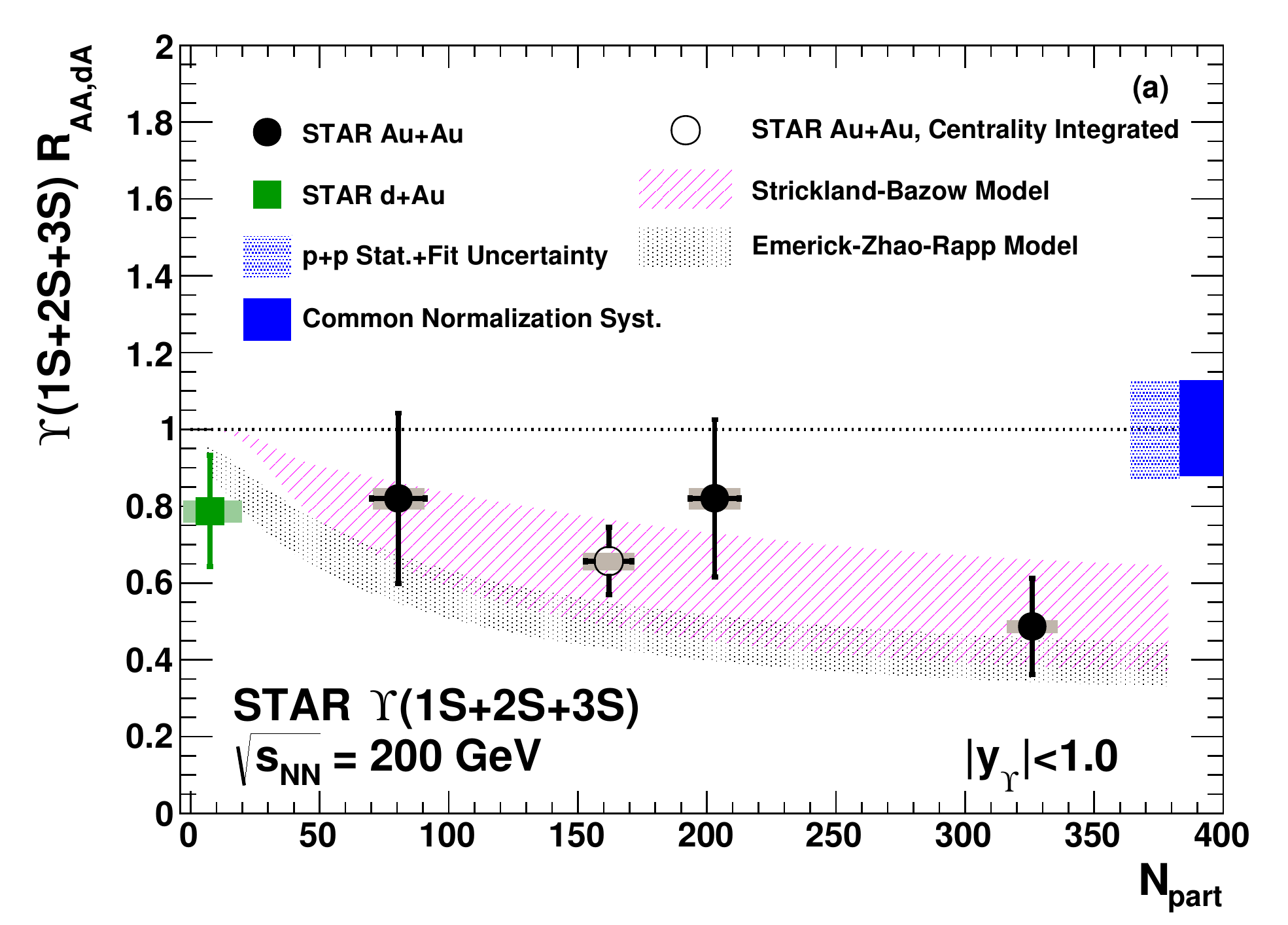} 
\includegraphics[width=0.40\textwidth, trim=1.8cm 0cm 0cm 0cm]{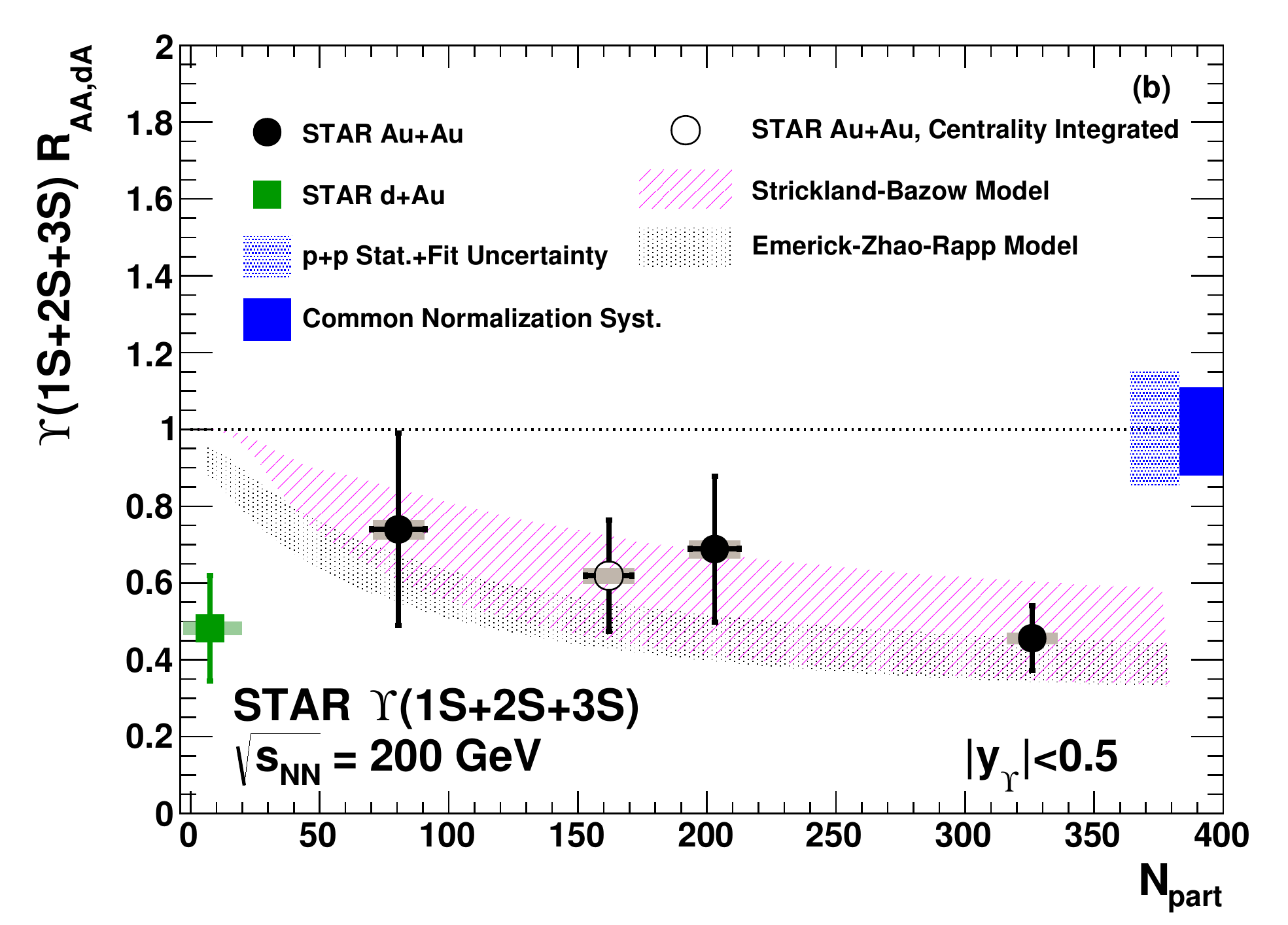}
\includegraphics[width=0.40\textwidth, trim=1.8cm 0cm 0cm 0cm]{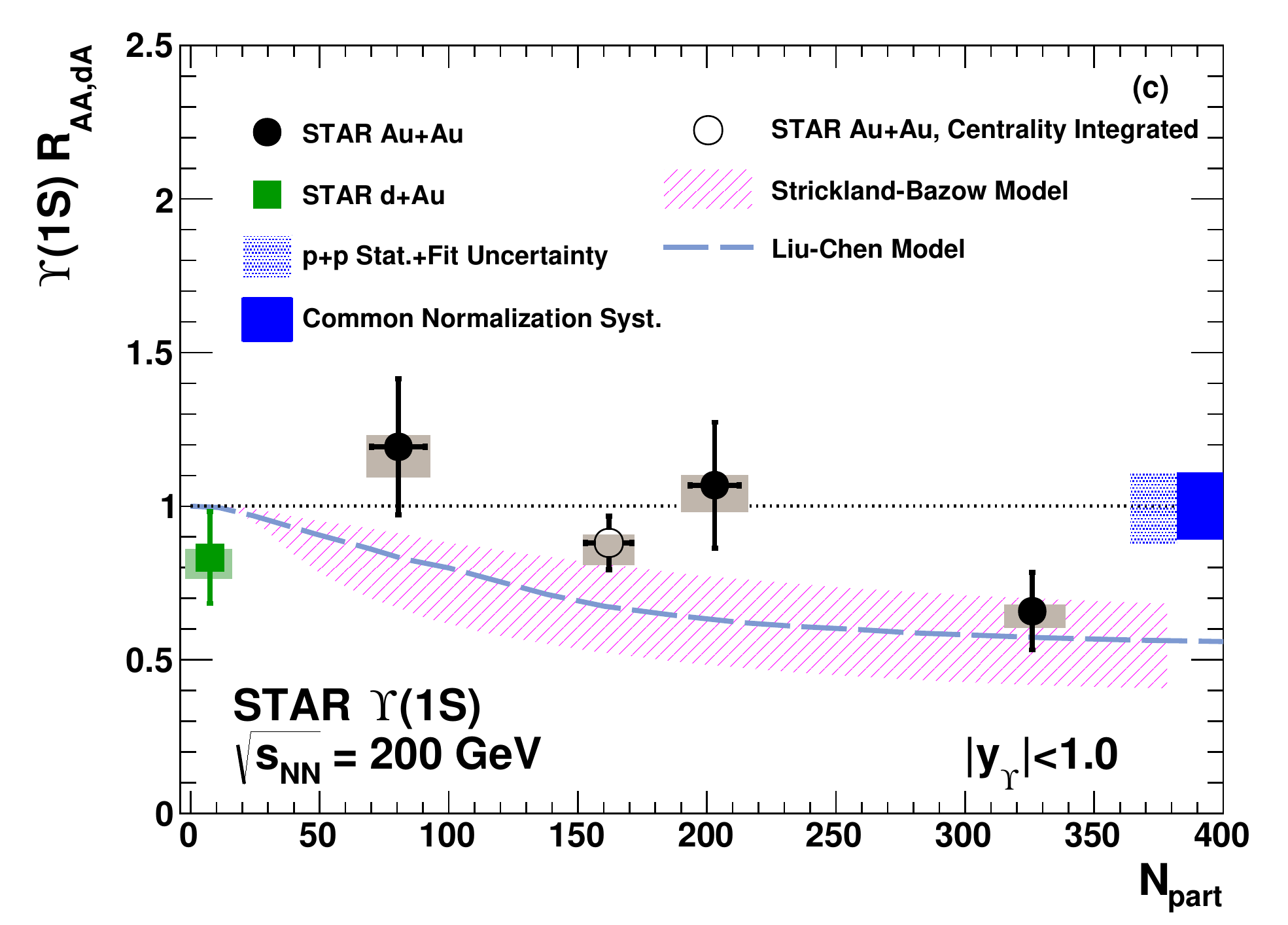} 
\caption{(Color online) Nuclear modification factor for \upsi(1S+2S+3S), in $|y|<1.0$ (a) and in $|y|<0.5$ (b), and \upsi(1S) in $|y|<1.0$ (c), in \dAu\ (green square) and \AuAu\ (black circles) collisions as a function of $N_{part}$. The boxes around unity show the statistical (shaded) and systematic (filled) uncertainty from the \pp\ measurement. The gray bands around the data points are the systematic uncertainties. The data are compared to calculations from Refs.~\cite{Strickland:2011aa,Emerick:2011xu,Liu:2010ej}.}
\label{fig:RAA}
\end{figure}
This suppression is quantified in Fig.~\ref{fig:RAA}, 
which displays the nuclear modification factor, \RAA,  plotted as a function of \Npart\ with the 0-10\% most-central 
collisions corresponding to $\langle\Npart\rangle=326\pm4$.
Figure~\ref{fig:RAA}(a) shows the data \textcolor{black}{for all three states} in the rapidity range $|y|<1$, 
while Fig.~\ref{fig:RAA}(b)
is for the narrower $|y|<0.5$ range. 
Figure~\ref{fig:RAA}(c) shows \RAA\ and \RdAu\ for the ground state \upsione\ 
in the range $|y|<1.0$. The data confirm that bottomonia are indeed suppressed in \dAu\ and in \AuAu\ collisions.   
For \dAu\ collisions, we find 
\textcolor{black}{$\RdAu(1S+2S+3S) = 
0.79 \pm 0.14\ (\dAu\ \mathrm{stat.}) \pm 0.10\ (\pp\ \mathrm{stat.}) 
\pm 0.03\ (\dAu\ \mathrm{syst.}) \pm 0.09 (\pp\ \mathrm{syst.})$} 
in the range $|y|<1$.  
We use a total inelastic cross section 
for \pp\ collisions of 42 mb, for \dAu\ collisions of 2.2 b, 
and $\langle N_{coll} \rangle =7.5 \pm 0.4$ for calculating \RdAu.
In the same rapidity range and for the 0-10\% most-central \AuAu\ collisions, 
we find 
\textcolor{black}{$R_{AA}(1S+2S+3S) = 0.49 \pm 0.13\ (\AuAu\ \mathrm{stat.}) \pm 0.07\ (\pp\ \mathrm{stat.}) 
\pm 0.02\ (\AuAu\ \mathrm{syst.}) \pm 0.06\ (\pp\ \mathrm{syst.})$,
which is $\approx 4.5\sigma$ away from unity.}
The results are summarized in Tab.~\ref{tab:RAA}.  

In the narrower rapidity range (Fig.~\ref{fig:RAA}(b)),
we see an indication of a lower \RdAu\ as discussed earlier.  
Our data and the E772 data show
a larger suppression at $y\sim0$ or $x_F\sim0$ than that expected from shadowing. 
The level of suppression we observe for $|y|<0.5$ stays approximately constant from \dAu\ up to central \AuAu\ collisions. 
This suggests that suppression in \dAu\ in this kinematic range needs to be understood
before interpreting the suppression in \AuAu.

For \dAu\ collisions we find 
\textcolor{black}{\RdAu(1S) $= 0.83 \pm 0.15\ (\dAu\ \mathrm{\ stat.}) 
    \pm 0.11\ (\pp\ \mathrm{\ stat.}) ^{+0.03}_{-0.07}\ (\dAu\ \mathrm{syst.}) 
\pm 0.10 (\pp\ \mathrm{syst.})$} in the range $|y|<1.0$.
For the 0-10\% most-central collisions we find 
\textcolor{black}{$R_{AA}(1S) = 0.66 \pm 0.13\ (\textcolor{black}{\AuAu} \mathrm{\ stat.}) \pm 0.10\ (\pp\ \mathrm{\ stat.}) 
^{+0.02}_{-0.05}\ (\textcolor{black}{\AuAu} \mathrm{\ syst.}) \pm 0.08\ (\pp\ \mathrm{syst.})$.}
Similar suppression is found by CMS in \PbPb\ collisions (\RAA(1S)$\approx0.45$ at similar \Npart)
\cite{Chatrchyan:2011cms, Chatrchyan:2012np, Chatrchyan:2012lxa}.
\textcolor{black}{We observe
the nuclear modification factor for the \upsione\ as a function of \Npart\ to be consistent with unity in \dAu\ 
through mid-central \AuAu\ collisions (see Fig.~\ref{fig:RAA}c).
In the most central \AuAu\ collisions, we see an indication of suppression of the \upsione\ at the $2.7\sigma$ level.
In the context of suppression of the excited states, if the feed-down fraction remains $\sim49\%$ as measured at higher energies and high-\pT\, it is possible that an \RAA(1S) as low as 0.51 could be due solely to suppression of the excited states \cite{Affolder:1999wm}.}

\textcolor{black}{One can relate the \RAA\ of the combined states to that of the ground state via the equation $\RAA(1S+2S+3S) = \RAA(1S) \textcolor{black}{\times} (1+ N_{AA}(2S+3S)/N_{AA}(1S))/(1+N_{pp}(2S+3S)/N_{pp}(1S))$.
    The ratio of the excited states to the ground state can be obtained from measurements by CMS and Fermilab experiments \cite{Khachatryan:2010zg,Moreno:1990sf} and alternatively from combining theoretical calculations \cite{Frawley:2008kk} with measured branching ratios from the PDG \cite{pdg}.
    In the case where $N_{AA}(2S+3S)=0$, \RAA(1S+2S+3S) $\approx \RAA(1S) \times 0.7$.}
This is consistent with our observed $\RAA$ values, and can also be inferred 
by examining the mass range 10--11 \gevcc\ in Fig.~\ref{fig:InvariantMassAuAuCentBins}, 
where no significant 2S or 3S signals are seen. 

\begin{figure}[t]
  \includegraphics[width=0.48\textwidth]{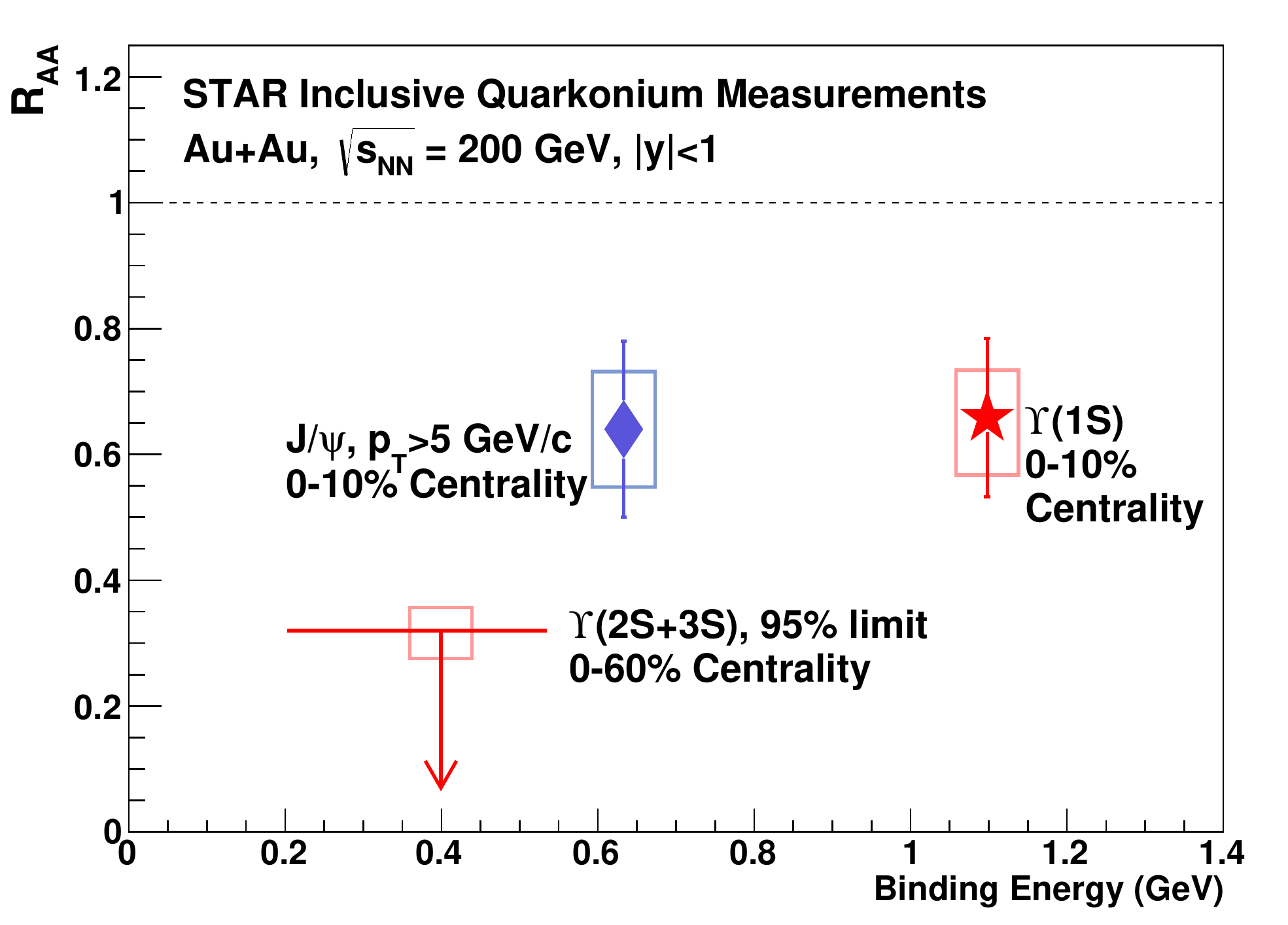}
  \caption{(Color online) Nuclear modification factor of quarkonium states as a function of binding energy as measured by STAR.
  The horizontal line of the $\upsi$(2S+3S) upper limit spans the range from the 3S to 2S binding energy; the arrow is placed at the weighted average of the binding energies.
  The high-$p_{T}$ J/$\psi$ results are from Ref.~\cite{Adamczyk:2012ey}.}
  \label{fig:BindingEnergy}
\end{figure}

By applying the methods described in \cite{Rolke}, we can calculate an upper limit on the \RAA\ of the combined 2S and 3S states.
Using the fit to Drell-Yan and \bbbar\ (dashed, green curve) as the background, we find an upper limit of \textcolor{black}{29} signal counts with 95\% confidence in the mass range 10--11 \gevcc\ for 0--60\% centrality collisions.
To transform this upper limit into an upper limit on \RAA(2S+3S), we assumed that the 
purity of excited states in this mass range
is the same as in the \pp\ case. 
While the excited states are likely more suppressed than the ground state in the \AuAu\ case, 
using the \pp\ purity gives us an upper limit in the \AuAu\ purity which can be used to calculate 
an upper limit on the \RAA.
The 2S+3S cross section in \pp\ was extracted from the full cross section, assuming the purity 
can be obtained based on the PDG branching ratios \cite{pdg} and the relative production
cross sections of the three states.
In the centrality range of 0--60\%, we thus obtain a 95\%-confidence upper limit of \textcolor{black}{\RAA(2S+3S) $< 0.32$ (see Fig.~\ref{fig:BindingEnergy})}.

Our data are also compared to model calculations incorporating
hot-nuclear-matter effects for \AuAu ~\cite{Strickland:2011aa, Emerick:2011xu, Liu:2010ej}.
These aim to incorporate lattice-QCD results pertinent to screening and 
broadening of bottomonium and to model the dynamical propagation 
of the $\Upsilon$ meson in the colored medium. 
Both models are in agreement with the level
of suppression seen in \AuAu. The model proposed by Emerick, Zhao, and Rapp (EZR), Ref.~\cite{Emerick:2011xu},
includes possible CNM effects,
modeled as an absorption cross section of up to 3 mb which 
can account for a value of \RAA\ as low as 0.7.
In this model the additional suppression to bring \RAA\ 
down to $\approx 0.5$ is due to hot-nuclear-matter effects.
The calculation by Liu et al.~\cite{Liu:2010ej} in Fig.~\ref{fig:RAA}(c) is for the inclusive \upsione\ \RAA, using
the internal energy as the heavy-quark potential and
an initial temperature of the fireball of $T=340$ MeV, which given the input from lattice QCD results, is
not hot enough to melt the directly produced \upsione.  Hence,
the suppression is mostly driven by the dissociation of the excited states (both the S-states and the P-states). 
The initial temperature
used in the EZR model is 330 MeV (with a formation time of 0.6 fm/c). 
The temperatures of the QGP needed in Strickland's model, Ref.~\cite{Strickland:2011aa},
are in the range 428 -- 442 MeV.  However, it should be noted that neither the Strickland model, nor the
calculation from Liu et al.~include any CNM effects.

\begin{figure}[h!]
\includegraphics[width=0.48\textwidth]{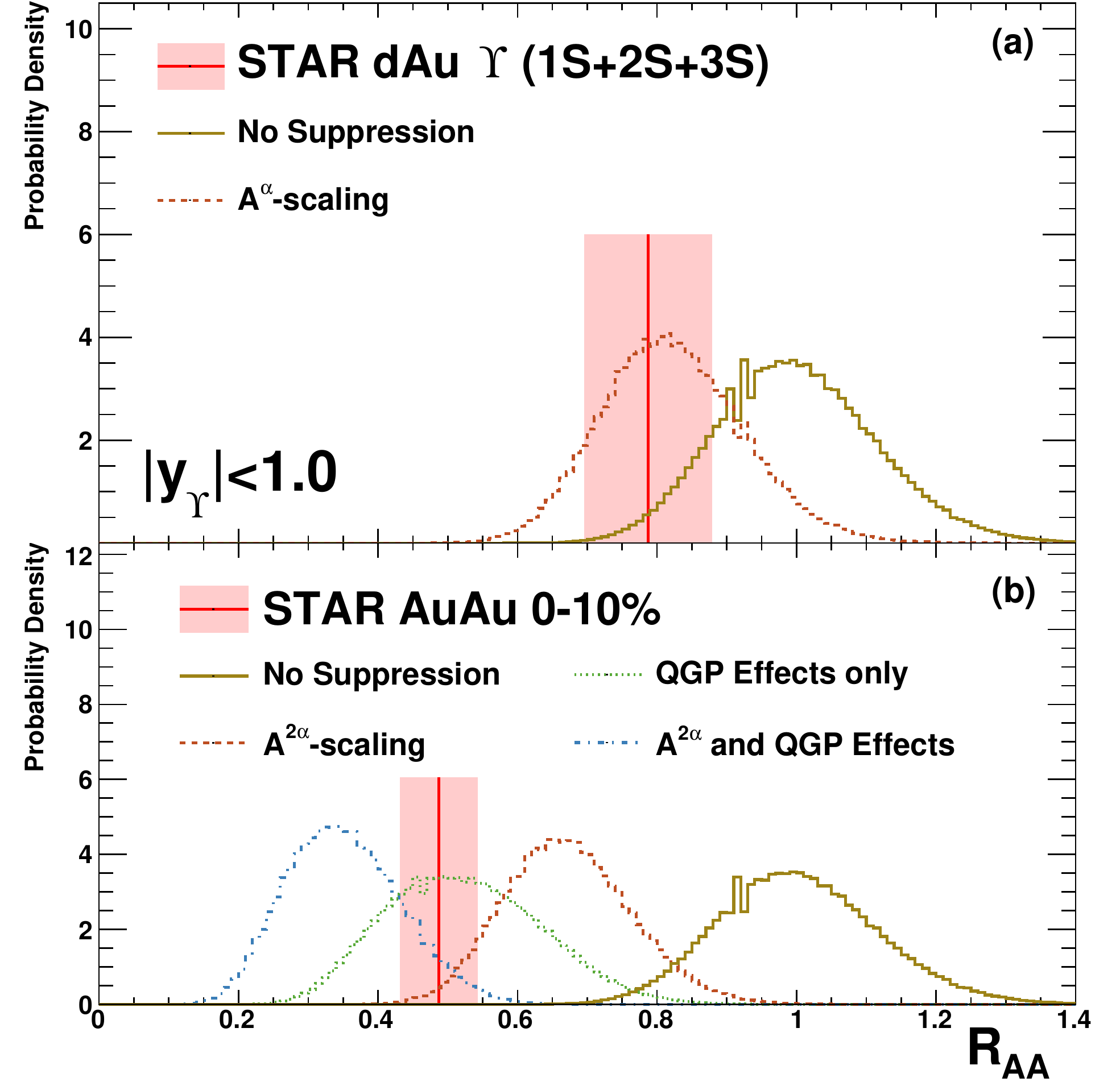}
\includegraphics[width=0.48\textwidth]{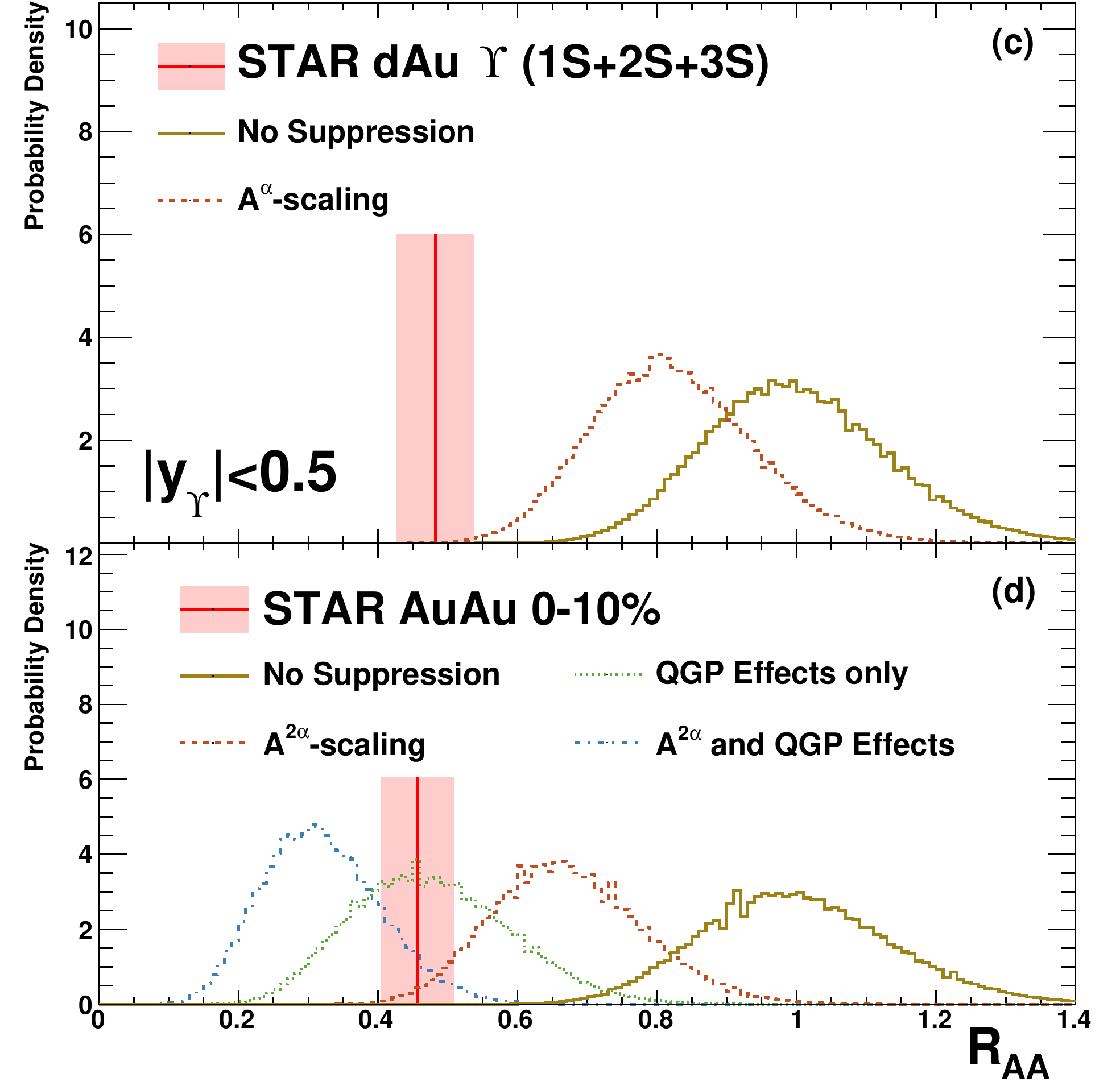}
\caption{(Color online) Summary of the results of four different pseudoexperiments: No Suppression (solid gold), CMN effects only (dashed red), QGP effects only (dotted green), and both CMN and QGP effects (dot-dashed blue). \textcolor{black}{We show our results for two systems and two rapidity ranges: (a)~\dAu\ $|y|<1.0$, (b)~\AuAu\ $|y|<1.0$, (c)~\dAu\ $|y|<0.5$, (d)~\AuAu\ $|y|<0.5$.} Our data is shown as a red vertical line with systematics shown by the pink box. The QGP effects are modeled in \cite{Strickland:2011aa}.}
\label{fig:MC_0_10}
\end{figure}

\textcolor{black}{Considering} two possible sources of suppression, CNM and QGP effects, we used a Monte Carlo pseudoexperiment to compare our results to different possible sources of suppression. 
We investigated four possible scenarios: (1) No suppression compared to \pp; (2) Suppression due to CNM effects only;
(3) QGP suppression only; (4) Suppression from both CNM and QGP effects. We simulated $\Upsilon$ production in \pp, \dAu, and \AuAu\ collisions via a Poisson generator. 
CNM effects were included via the suppression parametrization used by E772 \cite{Alde:1991sw} and presented in
Fig.~\ref{fig:E772}(a). 
We used the predictions from the Strickland model \cite{Strickland:2011aa} to estimate suppression from QGP effects. 
For scenario (4), the expected suppression is simply taken to be the product of the suppression from scenario (2) and scenario (3).
For this pseudoexperiment we assumed a flat prior 
based on the allowed \RAA\ given in Strickland-Bazow \cite{Strickland:2011aa},
depicted as the band for this calculation in Fig.~\ref{fig:RAA},
stemming from the choice of $1<4\pi\eta/S<3$.

A summary of the pseudoexperiment results is shown in Fig.~\ref{fig:MC_0_10}. 
Panel (a) shows our result for \RdAu\ in the range $|y|<1.0$ compared to scenarios (1)
and (2), shown as the solid line and dotted histogram, respectively.  
The \textcolor{black}{`no-cold-suppression-scenarios' (1 and 3)} are excluded while the CNM effects from E772 parameterization
are consistent with our observation.
Panel (b) shows \RAA\ for the most-central \AuAu\ bin in the range $|y|<1.0$.
By comparing the results of the pseudoexperiments with our measurements, 
we are able to exclude scenario (1) at a $\sim 5\sigma$ confidence level. 
Finally, we see that hypothesis (4) (dot-dashed curve), including both hot and cold nuclear effects,
is consistent with our measurements \textcolor{black}{when both the \dAu\ and \AuAu\ results are taken into account.}

We repeated this procedure for the rapidity range $|y|<0.5$. 
The results are shown in Figs.~\ref{fig:MC_0_10} (c) and (d). 
In the mid-rapidity range we find a larger amount of suppression in \dAu\ 
than what we observe in the range $|y|<1.0$. 
Furthermore, \RdAu\ is comparable to \RAA\ in 0-10\% for this rapidity range.
This could indicate that suppression of bottomonium already occurs in \dAu\ collisions.
\textcolor{black}{However, given the uncertainties in our current results, no particular model of \upsi\ suppression in \dAu\ is favored.}
Hence, further investigation of cold-nuclear-matter effects on \upsi\ production is highly warranted.
The suppression effects seen in \dAu, which are not explained by the models discussed here,
still need to be understood before the \AuAu\ results can be fully interpreted.

\section{Conclusions}
\label{sec:Conclusions}

In conclusion we studied  \upsi(1S+2S+3S) 
production in \pp, \dAu, and \AuAu\  collisions at \sqrts=200 GeV. 
We measured 
the cross section in \pp\ collisions to be 
\textcolor{black}{$B_{ee}\times d\sigma/dy |_{|y|<1}= 61 \pm 8 (\mathrm{stat.+fit}) ^{+13}_{-12} (\mathrm{syst.})$ pb}
and find it to be consistent within errors with NLO calculations.  
The cross section in \dAu\ collisions is found to be
\textcolor{black}{$B_{ee}\times d\sigma/dy |_{|y|<1}= 19 \pm 3 (\mathrm{stat.+fit}) \pm3 (\mathrm{syst}.)$ nb.}
We obtain a nuclear modification factor in this rapidity region ($|y|<1$) of
\textcolor{black}{$\RdAu(1S+2S+3S) = 
0.79 \pm 0.14\ (\dAu\ \mathrm{stat.}) \pm 0.10\ (\pp\ \mathrm{stat.}) 
\pm 0.03\ (\dAu\ \mathrm{syst.}) \pm 0.09 (\pp\ \mathrm{syst.})$.}
Models of \upsi\ production in cold nuclear matter, which include 
shadowing and initial-state partonic  energy loss, 
\textcolor{black}{are consistent with} the cross-sections we observe in our \dAu\ data. 
\textcolor{black}{Higher statistics \dAu\ data are required to further investigate the $3\sigma$ deviation we observe at $|y|<0.5$.}
We measured the \upsi(1S+2S+3S) nuclear 
modification factor in \AuAu\ collisions at \sqrtsNN=200 GeV as a function of centrality. 
In the range $|y|<1$ and in 0-10\% most-central 
collisions we find 
\textcolor{black}{$R_{AA}(1S+2S+3S) = 0.49 \pm 0.13\ (\AuAu\ \mathrm{stat.}) \pm 0.07\ (\pp\ \mathrm{stat.}) 
\pm 0.02\ (\AuAu\ \mathrm{syst.}) \pm 0.06\ (\pp\ \mathrm{syst.})$}, 
indicating additional \upsi\ suppression in hot nuclear matter compared to cold nuclear matter. 
In 0-60\% centrality we find a 95\%-confidence upper 
limit on the nuclear modification of the excited states of \textcolor{black}{$R_{AA}$(2S+3S)$ < 0.32$}.
Calculations of the centrality dependence of \upsi\ $R_{AA}$ 
using models based on lattice QCD calculations of bottomonium melting 
in a hot medium are found to be consistent with our data.  
Therefore, the suppression seen in central \AuAu\ collisions is indicative of the presence of deconfined matter in heavy-ion collisions.
It would be desirable to have a higher statistics \dAu\ dataset in order to strengthen the conclusions regarding cold-nuclear modifications to \upsi\ production before a stronger connection between parton deconfinement, Debye screening, and the observed \upsi\ suppression in \AuAu\ can be made.

\section{Acknowledgements}
\label{sec:Acknowledgements}
We thank R. Vogt, M. Strickland, R. Rapp, F. Arleo, and J.P. Lansberg for providing us calculations in the STAR kinematic regions. 
We thank the RHIC Operations Group and RCF at BNL, the NERSC Center at LBNL, the KISTI Center in Korea, and the Open Science Grid consortium for providing resources and support. This work was supported in part by the Offices of NP and HEP within the U.S. DOE Office of Science, the U.S. NSF, CNRS/IN2P3, FAPESP-CNPq of Brazil,  the Ministry of Education and Science of the Russian Federation, NNSFC, CAS, MoST and MoE of China, the Korean Research Foundation, GA and MSMT of the Czech Republic, FIAS of Germany, DAE, DST, and CSIR of India, the National Science Centre of Poland, National Research Foundation (NRF-2012004024), the Ministry of Science, Education and Sports of the Republic of Croatia, and RosAtom of Russia.

\end{document}